\documentclass[preprint,12pt]{elsarticle}
\usepackage{mdframed}

\usepackage{amssymb}

\journal{a journal of physics}

\usepackage[hyphens]{url} 
\usepackage{subfig} 
\usepackage{rotating} 
\usepackage{comment}
\usepackage{tikz}
\usepackage{float}
\usepackage{verbatim}
\usepackage{hyperref}

\hypersetup{
    colorlinks=true,
    linkcolor=blue,
    citecolor=blue,
    filecolor=blue,
    urlcolor=blue,
}

\usepackage[hyphens]{url} 
\usepackage{subfig} 

\usepackage{multicol}
\usepackage[utf8]{inputenc}
\usepackage{csquotes}

\usepackage{caption}
\usepackage{amsmath}
\usepackage{changepage}
\usepackage{courier}
\usepackage{colortbl} 

\usepackage{booktabs} 

\usepackage{float}
\usepackage{placeins} 
\usepackage{geometry} 
\usepackage{verbatim}
\usepackage{listings} 
\usepackage{listingsutf8}
\usepackage{booktabs} 

\usepackage{tabularx}

\usepackage{graphicx}
\usepackage{pdflscape}
\usepackage{adjustbox}

\graphicspath{{./imgs/}}

\usepackage[labelfont=bf]{caption} 

\usepackage[labelsep=period]{caption}

\begin{document}

\begin{frontmatter}





    \title{Questions of science: chatting with ChatGPT about complex systems}


	 \author[uff]{Nuno Crokidakis}

	 \author[uff,INCT]{Marcio Argollo de Menezes}

	 \author[eco,INCT,LAMFO]{Daniel O. Cajueiro}

	 \address[uff]{Instituto de F\'isica, Universidade Federal Fluminense, Niter\'oi/RJ, Brazil}

	 \address[INCT]{National Institute of Science and Technology for Complex Systems (INCT-SC), Brazil.}

	 \address[eco]{Department of Economics, University of Brasilia, Campus Universit\'ario Darcy Ribeiro - FACE, Brasilia, DF 70910-900.}

	 \address[LAMFO]{Machine Learning Laboratory in Finance and Organizations (LAMFO), Universidade de Bras\'{i}lia (UnB), 70910-900, Bras\'{i}lia, Brazil.}


	 \cortext[cor1]{Corresponding author: danielcajueiro@gmail.com }

	\begin{abstract}
We present an overview of the complex systems field using ChatGPT as a representation of the community's understanding. ChatGPT has learned language patterns and styles from a large dataset of internet texts, allowing it to provide answers that reflect common opinions, ideas, and language patterns found in the community. Our exploration covers both teaching and learning, and research topics. We recognize the value of ChatGPT as a source for the community's ideas.

	\end{abstract}


\end{frontmatter}


\clearpage

\tableofcontents

\clearpage

\section{Introduction}

ChatGPT and other transformers-based large language models (LLMs) have demonstrated remarkable abilities to perform tasks such as completion tasks, question answering, reading comprehension and translation  that were once thought to be exclusive to humans \citep{devlin2018bert,zhang2020pegasus,raffel2019exploring,fedus2021switch,lewis2019bart,radford2018improving,radford2019language,brown2020language}. 

We are currently in a great era for researchers and scientists studying and developing in the field of complex systems. Half of the physics Nobel prize of 2021 was awarded to the physicist Giorgio Parisi  for his contributions to the theory of complex systems \citep{Cugliandolo_2023} and the other half to two  meteorologists  Syukuro Manabe and Klaus Hasselmann to the modeling of the Earth’s climate \citep{gupta2022perspectives}.  Parisi  has made significant contributions to the literature on complex systems, including areas such as spin glass \citep{Parisi1979,Parisi1980,Parisi_1980order}, stochastic resonance  \citep{Benzi1982}, surface growth \citep{Kardar1986}, multifractality \citep{frisch1985fully}, and bird flocking \citep{Ballerini2008}.  Manabe showed that there is a relation between the increase in global temperature and the level of CO${_2}$ in the atmosphere \citep{manabe1967thermal}.  Hasselmann developed a model that establishes a connection between weather and climate, implying that climate models can be reliable despite the unpredictability and complexity of weather patterns \citep{hasselmann1976stochastic,frankignoul1977stochastic}. The diversity of research topics among the award recipients highlights the multidisciplinary nature of the field. In addition, the applied dimension of the contributions stresses the importance of applied contributions to the field. In particular, it is worth mentioning that the research of field of complex systems has developed  multiple frameworks, such as chaos theory \citep{lorenz1963deterministic}, fractal theory \citep{mandelbrot1982fractal}, complex networks \citep{barabasi1999emergence,watts1998collective}, and agent-based models \citep{bonabeau2002agent}, to model and solve important practical problems in diverse fields.

OpenAI has just now introduced the ChatGPT, a large language model tool trained, to aid humans in a variate of human tasks including question answering, text edition and has been used as a great information provider \citep{brown2020language,ouyang2022training}. ChatGPT has been trained on a large dataset of internet texts, which means that it has learned to mimic the patterns and styles of language used by many different people online. When you talk to ChatGPT, it responds based on what it has learned from that dataset, so its answers can be thought of as representing the average opinions, ideas, and language patterns that are commonly found on the internet. 

In this paper, we overview different aspects of the field of complex systems by means of chats with ChatGPT. We split this paper into two parts. The first part explores issues related to teaching and learning, where ChatGPT may work as a teacher assistant. The second part explores issues related to research, where, in this case, ChatGPT works like a research assistant.  Besides exploring different concepts from different points of view in the field of complex systems, we also recognize that ChatGPT can be a valuable resource for teachers and scientists. 


\section{Large language models and ChatGPT}

The precursors of the large language models that we know today are the classical and sparse ones known as $n$-grams models \citep{shannon1948mathematical,manning1999foundations}. In Natural Language Processing (NLP), $n$-gram models are probabilistic models that aim to predict the next word in a text using a limited amount of previous context. The value of $n$ indicates the number of words that are used to establish this context. Mathematically speaking, the goal of an $n$-gram model is to estimate the probability $P(t_n|t_1,\cdots,t_{n-1})$ of a particular token (word), $t_n$, occurring based on the history of words that have come before it, specifically $t_1$ through $t_{n-1}$. We estimate these probabilities by counting the number of times that a sequence of tokens arise in a large corpus. We may consider, for instance, that we intend to estimate the probability of the word ``mountains" after the history ``These mist covered" that comes from the first sentence ``These mist covered mountains
are a home now for me" of the song ``Brothers in Arms" of the British band Dire Straits.  In order to estimate this probability, we have to count the number of times the sequence of tokens $t_1$=``These", $t_2$= ``mist", $t_3$= ``covered" arise before $t_4$=``mountains". Thus, we may estimate $P(t_4|t_1,t_2,t_3)$ as $\frac{P(t_1,t_2,t_3,t_4)}{P(t_1,t_2,t_3)}$, where $P(t_1,t_2,\cdots,t_n)$ is the frequency of the $n$-gram $t_1,\cdots,t_n$ in the corpus.

Most popular language models today are based neural network models. Although there are many different types of neural network models, all share some ingredients with the Multi-Layer Perceptron (MLP)\citep{rumelhart1986learning}. 
An MLP consists of multiple layers of interconnected nodes, called ``neurons", where each neuron receives input from the neurons in the previous layer and produces an output that is sent to the neurons in the next layer. The input to the first layer is typically a vector of features and the output of the last layer is the desired output. Each neuron in an MLP computes a linear combination of its inputs and applies a nonlinear function, called ``activation function", to the result. The weights on the inputs determine the strength of the connections between neurons. The nonlinear activation function is necessary to ensure a neural network is able to approximate any continuous function \citep{cybenko1989approximation}. We may describe mathematically the outputs of the neurons of layer $l$ of MLP as

\begin{equation}y_{j}^{(l)}=h^{(l)} \left(\sum_{i=1}^{M_{l-1}} w_{ij}^{(l)}y_{i}^{(l-1)} + 
w_{0j}^{(l)}\right), \; j=1,\cdots, n_{l},\end{equation}
\noindent where $n_l$ is the number of neurons of layer $l$, $w_{ij}$ is the weight associated with the input $i$ of the (previous) layer $l-1$ and the neuron $j$ of layer $l$,  $y_{i}^{(l-1)}$ is the $i$th output of the (previous) layer $l-1$, $M_l$ is the number of neurons of layer $l$, and $h^l$ is the activation function of layer $l$\footnote{Common types of activation functions are tanh, sigmoide and Relu.}. We call the term  \(a_{j}^{(l)}=\sum_{i=1}^{M_{l-1}} w_{ij}^{(l)}y_{i}^{(l-1)} + 
w_{0j}^{(l)}\) activation of the neuron \(j\)\footnote{It is also denoted as \(a_{j}^{(l)}=\sum_{i=0}^{M_{l-1}} w_{ij}^{(l)}y_{i}^{(l-1)}\), where we associate the bias (constant) \(w_{0j}^{(l)}\) with the input \(y_{0}^{(l-1)}=1\).}. The input of the first layer is \(y_{i}^{0}=x_i\), which is actually the input of the neural network with \( M_0\) attributes.  The output of the neural network is 
 \(y_{i}^{nl}=y_{i}^{0}\), for \(i=1,\cdots M_{nl}\). Thus, the neural network has  \(M_{nl}=M_o\) outputs.

In order to estimate the weights (parameters) of a MLP, we need a dataset with a pairs of inputs and outputs. Since the outputs provide a reference for what the neural network should learn for each input, we call this kind of learning as {\it supervised learning}. 
We use the gradient descent (ascent) to minimize (maximize) the output square error (maximal likelihood associated to the problem).  Since we have only the errors (or the likelihoods) associated with the last layer, we use the chain rule of calculus to evaluate the gradients associated with the weights of the other layers. This is the so-called {\it backpropagation algorithm}.

A crucial development towards the creation of large language models has been the use of word embeddings to represent words. {\it Word embeddings} are vectors that represent words in high dimensional Euclidean spaces \citep{bengio2000neural}. The concept of word embeddings originates from the idea that the meaning of a word can be inferred from the context in which it appears. We may generate word embeddings using neural network models that learn to predict a word based on its surrounding context. One of the most common neural network models used to generate word embeddings is the Continuous Bag-of-Words (CBOW) model \citep{mikolov2013efficient}. The CBOW model consists of three layers: an input layer, a hidden layer, and an output layer. The input layer encodes the context words as one-hot vectors\footnote{A one-hot vector is a vector that has the same dimensionality as the vocabulary size and has exactly one element set to 1 and all other elements set to 0. The position of the 1 corresponds to the index of the word in the vocabulary.}, while the output layer predicts the probability distribution over the vocabulary for the target word. The hidden layer maps the input vectors to a lower-dimensional representation, which is the word embedding. The input layer has $C$ inputs (the size of the context used), the intermediate layer has size $N$ (the size of the word embedding vectors) and the output layer has $N_V$ outputs. The size of the hidden layer, as well as the number of context words used as input (denoted by $C$), are hyperparameters that we need to tune for each application. Thus, we have to estimate $C$ context matrices of size $N_V\times N$ and one output matrix of size $N\times N_V$.   
Note that the problem we solve here is very similar to the one considered by the $n$-grams models. Consider, for instance, the first sentence of the song ``All by myself" of Eric Carmen ``When I was young I never needed anyone". Suppose that we consider a context of size 6. Thus, we use the tokens $t_1$=``I" ,$t_2$=``was", $t_3$=``young ", $t_5$=``never", ,$t_6$=``needed" and ,$t_7$=``anyone" to predict ,$t_4$=``I".  
To generate the word embeddings, the model is trained on a large corpus of text in a supervised fashion. During training, the weights of the network are updated to minimize the difference between the predicted probability distribution and the actual distribution of the target word. The actual word embedding vector that is supposedly to represent a word is the average of the rows of the context matrices corresponding to the word and the corresponding column of the output matrix.

Transformers form the foundation of ChatGPT and other contemporary large language models.
They are neural network models that consist of three ingredients \citep{vaswani2017attention}: a layer that represent the words using word embedding vectors, a position encoding, and a mechanism of attention.  The position encoding informs the position of each token in the input sequence and it is added to the corresponding word embedding vectors. This is achieved through a sinusoidal function that creates a unique encoding for each position in the sequence. The self-attention mechanism allows the model to focus on different parts of the input sequence depending on the task at hand. Self-attention operates by representing the input sequence as a set of queries, keys, and values\footnote{These concepts come from the field of information retrieval. The query is a short piece of text, such as a keyword or phrase, that the user enters into a search engine. The key is the piece of information associated with the object that is being searched. The value is the object itself the user is looking for.}, that are matrices that should be learned. The resulting attention scores are then used to weight the values, producing an output that is a weighted sum of the values, with the weights determined by the attention scores. There is today a large list of transformers that have been used in successful NLP tasks such as  
Google’s BERT \citep{devlin2018bert}, PEGASUS \citep{zhang2020pegasus}, T5 \citep{raffel2019exploring}, and Switch \citep{fedus2021switch}, Facebook’s BART \citep{lewis2019bart}, and Open AI’s Generative Pre-Training (GPT) \citep{radford2018improving},
GPT-2 \citep{radford2019language}, and GPT-3 \citep{brown2020language}.

ChatGPT is built on GPT-3 \citep{brown2020language}, that were trained using a large amount of internet data. In general, we train language models to predict the next word, but not to deal with questions of users. A language model may invent facts, may present toxic behavior, or simply do not follow commands. In this case, we say that a language model is poorly aligned. In order to deal with that Open AI uses reinforcement learning from human feedback to align this model based on three steps, based on the feedback of 40 people hired to label the data  \citep{ouyang2022training}: (1) They collect a dataset of human-labeled comparisons between outputs of the models; (2) They train a reward model on this dataset to predict the output the humans prefer; (3) They use this reward model to fine tune the model using a proximal policy optimization algorithm \citep{schulman2017proximal} that is type of reinforcement learning algorithm\footnote{A reinforcement learning problem explores a situation where the objective is to map states to actions in order to maximize a numerical reward signal. A Policy gradient method is a reinforcement learning algorithm that learns a parametric representation of the policy and use gradient ascent to iteratively improve it. The proximal policy gradient algorithm constrains  the magnitude of the policy update at each iteration  reducing the risk of the policy diverging. } \citep{sutton2018reinforcement}.    

\section{Definition of a complex system}
\label{sec:definition_complex_systems}

The development of statistical mechanics by Maxwell, Boltzmann and Gibbs in the second half of the $19^{th}$ century introduced probabilistic reasoning as a new paradigm in Physics: given incomplete information regarding the initial state of a system, one cannot predict to which state it will evolve. Nevertheless, it is still possible to derive a probabilistic prediction of the final state reached by the system \citep{PARISI1999}, as long as the precise defintion of the initial state is exchanged for an ensemble of initial states.

Following G. Parisi \citep{PARISI1999}, we define a system as complex if its behavior crucially depends on the details of the system. They are often composed of many interacting elements with (emergent) properties that cannot be inferred from the property of a single element \citep{kadanoff, anderson1972}. Since the behavior of a complex system cannot be deduced without full knowledge of its basic elements and their interactions, one again resorts to the concept of probabilistic prediction to determine the probability distribution of behaviors given a set of systems, each with a given set of elements and interactions among them \citep{PARISI1999}. With such approach one can identify the most minimalist model which reproduces features observed in many distinct models falling into the same universality class \citep{kadanoff}.

We start this tour over the ideas, methods and contributions of the field of complex systems asking ChatGPT for the definition of a complex system. Although there is not a general accepted definition of a general complex system,  Figure \ref{fig:definition_complex_system} shows that ChatGPT provides a very good definition ``{\it A complex system is a system that is made up of a large number of interconnected and interdependent parts or components, where the behavior of the system as a whole is not easily predictable from the behavior of the individual components.}". It also highlights several ingredients presented in these systems and the fields of application. 

\begin{figure}[H]
\centering
\fbox{\begin{minipage}{36em}
\small
{\bf How do you define a complex system?
} 

\vspace{5mm}

 {\bf ChatGPT}: A complex system is a system that is made up of a large number of interconnected and interdependent parts or components, where the behavior of the system as a whole is not easily predictable from the behavior of the individual components. Complex systems are characterized by non-linear interactions, emergence of new behaviors or properties, and sensitivity to initial conditions and perturbations. Examples of complex systems include the human brain, ecosystems, the economy, traffic networks, and social systems. The study of complex systems involves interdisciplinary approaches from various fields including physics, biology, mathematics, computer science, and sociology, among others.

\end{minipage}}
\caption{Definition of a complex system.}
\label{fig:definition_complex_system}
\end{figure}

\section{Teaching and learning}

There are often numerous intriguing questions that arise when we seek to teach or learn a new subject. In this section, we explore these questions by utilizing ChatGPT as a proxy for the answers that can be found within the complex systems community. Section \ref{sec:teaching_topic} investigates which topics should we choose for a course. Section \ref{sec:teaching_new_topic}  explores the impact of a given topic in a course. Section \ref{teaching_sources} seeks for interesting sources about a subject. Section \ref{sec:same_concept_different_level} asks how to present the same concept in different levels. Section \ref{sec:teaching_connection} looks for connections between different concepts. While Section \ref{sec:teaching_learning_activities} requests some learning activities, Section \ref{sec:teaching_lab} searches for practical laboratory experiments.  In Section \ref{sec:teaching_code}, we focus on the issues related to coding.

\subsection{Topics to cover in a course}
\label{sec:teaching_topic}

Courses about complex systems are not very common regular courses in Physics Departments, although there are good examples, such as the one at the Center for the Study of Complex Systems at the University of Michigan\footnote{https://lsa.umich.edu/cscs/graduate-students/graduate-courses.html}, the one at the Complexity Sciences Center at the University of California, Davis\footnote{https://csc.ucdavis.edu/Courses.html}, the one at the University of Bologna\footnote{https://www.unibo.it/en/teaching/course-unit-catalogue/course-unit/2022/433619}, 
the one at the Abdus Salam International Centre for Theoretical Physics\footnote{https://www.ictp.it/opportunity/international-master-physics-complex-systems} and 
the one at the Politecnico di Torino\footnote{http://www.pcs.polito.it/presentation}.

Given the vast array of topics that could be deemed essential for a course on complex systems, a key question arises: what should be selected for inclusion in the curriculum? Thus, we ask this to ChatGPT considering a course for  third-
year undergraduate physics students. Figure \ref{fig:topic_3rd_complex_systems} shows that these topics are aligned with relevant sources of the field \cite{miller2009complex,kwapien2012physical,charbonneau2017natural,thurner2018introduction}.  The topics it suggests are {\it Graph Theory and Network Analysis} \citep{albert2002statistical,newman2006structure,cohen2010complex,estrada2012structure,posfai2016network}, {\it Celular Automata} \citep{wolfram1983statistical,ilachinski2001cellular,schiff2011cellular}, {\it Agent-Based Modeling} \citep{bonabeau2002agent,helbing2012agent,de2014agent}, {\it Chaos Theory} \citep{lorenz1996essence,strogatz2018nonlinear}, {\it Self-Organized criticality} \citep{bak1988self,jensen1998self,turcotte1999self,christensen2005complexity,markovic2014power} and {\it Complexity and Information Theory} \citep{shannon1948mathematical,kullback1951information,rissanen2007information,beck2009generalised}. In the next question presented in Figure \ref{fig:topic_phd_complex_systems}, we ask about a complex systems course for PhD students. The big differences between this answer and the previous one are that the descriptions of the topics are more advanced and ChatGPT also suggests the inclusion of machine learning methods \citep{jordan2015machine,silva2016machine,Tang2020introduction} that have become prominent in different investigations of complex systems. 

In Figure \ref{fig:prerequisites_course}, we  ask ChatGpt to present the necessary prerequisites for a complex systems course for PhD students. Its answer covers basic topics of mathematics, statistics, systems, computer programming and some topics that a student of physics could have learned in an undergraduate course of complex systems.

\begin{figure}[H] \centering \fbox{\begin{minipage}{36em}
\small
{\bf What are the recommended topics to cover in a complex systems course for third-year undergraduate physics students? Considering the content's level of difficulty and relevance, what would be the most suitable sequence for these topics?
} 

\vspace{5mm}

 {\bf ChatGPT}: A complex systems course for third-year undergraduate physics students should cover a range of topics that provide a comprehensive understanding of complex systems and their behavior. The following topics are recommended for such a course:

Introduction to complex systems: This should cover the basics of complex systems and their properties, including emergence, self-organization, and non-linearity.

Graph theory and network analysis: This topic should introduce students to graph theory and its applications in complex systems. It should cover topics such as random graphs, small-world networks, scale-free networks, and community detection.

Cellular automata: This topic should introduce students to cellular automata and their applications in modeling complex systems. It should cover topics such as the Game of Life, the forest fire model, and the Ising model.

Agent-based modeling: This topic should introduce students to agent-based modeling and its applications in complex systems. It should cover topics such as the Schelling model, the Axelrod model, and the Swarm model.

Chaos theory: This topic should introduce students to chaos theory and its applications in complex systems. It should cover topics such as the logistic map, the Lorenz system, and the Mandelbrot set.

Self-organized criticality: This topic should introduce students to self-organized criticality and its applications in complex systems. It should cover topics such as sandpile models, earthquakes, and avalanches.

Complexity and information theory: This topic should introduce students to complexity and information theory and their applications in complex systems. It should cover topics such as entropy, mutual information, and the complexity-entropy causality plane.

In terms of the sequence of topics, it is recommended to start with an introduction to complex systems, followed by graph theory and network analysis, and then move on to cellular automata and agent-based modeling. Chaos theory and self-organized criticality can be covered next, followed by complexity and information theory. However, the sequence can be adjusted based on the instructor's preferences and the students' background.

\end{minipage}} 
\caption{Topics of a course of complex systems for a third-year undergraduate student of physics.}
\label{fig:topic_3rd_complex_systems} 

\end{figure}

\begin{figure}[H] \centering \fbox{\begin{minipage}{36em}
\small
{\bf What if the course is for PhD students?
} 

\vspace{5mm}

 {\bf ChatGPT}: 
If the complex systems course is intended for PhD students, the recommended topics could be more advanced and in-depth compared to a course designed for undergraduate students. The following topics are recommended for such a course:

Introduction to complex systems: This should cover advanced topics related to complex systems, including the philosophical and mathematical foundations of complexity science.

Network science: This topic should cover advanced topics in network science, including community detection, network motifs, network dynamics, and multilayer networks.

Nonlinear dynamics: This topic should cover advanced topics in nonlinear dynamics, including bifurcation theory, chaos theory, and attractors.

Information theory: This topic should cover advanced topics in information theory, including measures of complexity, entropy, and mutual information.

Agent-based modeling: This topic should cover advanced topics in agent-based modeling, including the design and implementation of complex systems models, calibration and validation of models, and sensitivity analysis.

Machine learning and data analytics: This topic should cover advanced topics in machine learning and data analytics, including clustering, classification, regression, and deep learning, and their applications in complex systems analysis.

Complex systems applications: This topic should cover advanced applications of complex systems in various domains, such as social systems, biological systems, and ecological systems.

In terms of the sequence of topics, it is recommended to start with an introduction to complex systems, followed by network science and nonlinear dynamics. Agent-based modeling and information theory can be covered next, followed by machine learning and data analytics. Finally, the course can conclude with advanced applications of complex systems in various domains. However, the sequence can be adjusted based on the instructor's preferences and the students' background. The level of difficulty and mathematical rigor of the course should be appropriately adjusted to suit the PhD students' training and research interests.
\end{minipage}} \caption{Topics of a course of complex systems for a PhD student of physics.} \label{fig:topic_phd_complex_systems} \end{figure}

\begin{figure}[H] \centering \fbox{\begin{minipage}{36em}
\small
{\bf What should be the prerequisites for this course?

} 

\vspace{5mm}

 {\bf ChatGPT}: The prerequisites for a complex systems course depend on the course's intended audience and the level of difficulty of the topics covered. However, some general prerequisites for a complex systems course are as follows:

Mathematics: A strong background in mathematics is essential for understanding complex systems. Students should have a good understanding of calculus, linear algebra, differential equations, probability theory, and statistics.

Programming: Students should have experience in programming, preferably in a language such as Python, MATLAB, or R, which are commonly used in complex systems modeling and analysis.

Physics: A basic understanding of physics, including mechanics, thermodynamics, and electromagnetism, is helpful in understanding the dynamics of complex systems.

Systems thinking: Students should have a basic understanding of systems thinking, including feedback loops, system boundaries, and emergent behavior.

Complex systems concepts: Familiarity with the basic concepts of complex systems, such as emergence, self-organization, and non-linearity, is helpful in understanding the course's content.

Network theory: Familiarity with basic network theory concepts, such as graph theory, is useful in understanding the dynamics of complex networks.

Statistics and data analysis: Familiarity with basic statistical concepts, such as hypothesis testing and regression analysis, and experience in data analysis are helpful in understanding complex systems' modeling and analysis.

The specific prerequisites for a complex systems course may vary depending on the course's content and the instructor's preferences. It is recommended to consult the course syllabus or the instructor for more detailed information on the prerequisites.

\end{minipage}} \caption{Prerequisites for a complex systems course in PhD level.} \label{fig:prerequisites_course}

 \end{figure}

\subsection{Introduction of a specific topic and the impact of it in the course}
\label{sec:teaching_new_topic}

In Section \ref{sec:teaching_topic}, we ask about a recommendation for topics for a course on complex systems and ChatGPT does not include, for instance, the topic on {\it phase transitions}. 
Phase transitions are ubiquitous in physics \cite{stanley1987introduction}, and specially in complex systems \cite{RevModPhys.80.1275}. So, now, in the present section, we ask about the possibility of including it.   In Figure \ref{fig:phase_transitions}, we show that ChatGPT agrees with the inclusion of this topic and it explains that phase transitions can aid students in understanding 
the collective behavior and in modeling critical phenomena in a variety of systems. Then, in Figure \ref{fig:phase_transitions_example}, we ask about a specific example of phase transitions we may consider in the course. ChatGPT suggests that the {\it percolation transition} \citep{stauffer1979scaling,essam1980percolation,bollobas2006percolation,stauffer2018introduction} that illustrates several fundamental concepts in the study of complex systems, including the idea of criticality and the emergence of collective behavior.

\begin{figure}[H] \centering \fbox{\begin{minipage}{36em}
\small
{\bf What do you think about introducing the theory of phase transitions in an introductory course of complex systems?
} 

\vspace{5mm}

 {\bf ChatGPT}: Introducing the theory of phase transitions in an introductory course on complex systems can be a valuable addition to the course material. The theory of phase transitions is a fundamental concept in the study of complex systems, and it has many applications across different fields, including physics, chemistry, materials science, and biology.

Phase transitions refer to the abrupt changes in the properties of a system as a result of small changes in external parameters such as temperature, pressure, or magnetic field. The behavior of complex systems can often be described by phase transitions, and understanding these transitions can provide insights into the collective behavior of the system.

Moreover, phase transitions can be used to model critical phenomena in a variety of systems, including social networks, ecological systems, and financial markets. Therefore, introducing the theory of phase transitions in an introductory course on complex systems can help students develop a deep understanding of the behavior of complex systems and prepare them for further study in related fields.

Overall, I believe that introducing the theory of phase transitions in an introductory course on complex systems can be a valuable addition to the course material and can help students develop a deeper understanding of the behavior of complex systems.
\end{minipage}} \caption{Inclusion of a topic of phase transitions in a course of complex systems.} \label{fig:phase_transitions} \end{figure}

\begin{figure}[H] \centering \fbox{\begin{minipage}{36em}
\small
{\bf  Do you suggest any specific example of phase transition we may use in a complex systems course? Consider the sake of understanding and simplicity.
} 

\vspace{5mm}

 {\bf ChatGPT}: Sure! One example of a phase transition in a complex system that is often used in introductory courses is the percolation transition. Percolation refers to the process by which a fluid or gas passes through a porous material, such as soil or rock. In the context of complex systems, percolation can also refer to the process by which a connected network of nodes or clusters emerges in a system.

The percolation transition occurs when there is a critical probability of connections between nodes in the system, above which a giant connected component emerges that spans the entire system. Below this critical probability, the connected component is small and fragmented. This transition is an example of a second-order phase transition, which means that it is characterized by a continuous change in the system's properties as the probability of connections changes.

The percolation transition can be used to illustrate several fundamental concepts in the study of complex systems, including the idea of criticality and the emergence of collective behavior. It is a simple and intuitive example that can help students understand the fundamental concepts of phase transitions in complex systems.

Moreover, percolation has applications in a wide range of fields, including physics, biology, and computer science. For example, it can be used to model the spread of diseases in a population, the flow of electrical current through a network, or the behavior of fluid in porous materials.

Overall, the percolation transition is a useful example to introduce the theory of phase transitions in an introductory course on complex systems. It is a simple and relatable example that can help students understand the fundamental concepts of criticality, emergence, and universality in complex systems.
\end{minipage}} 
\caption{Example about phase transitions that we may use in  a course of complex systems.} \label{fig:phase_transitions_example}

\end{figure}

\subsection{Sources}
\label{teaching_sources}

Another important dimension of teaching or self-learning is the choice of good sources. We ask, in particular, in this section about possible sources for a possible course in  {\it complex systems} in a PhD level. ChatGPT answers this question providing the following references: {\it Networks, Crowds, and Markets: Reasoning About a Highly Connected World} \citep{easley2010networks}, {Nonlinear Dynamics and Chaos: With Applications to Physics, Biology, Chemistry, and Engineering} \citep{strogatz2018nonlinear}, {\it Introduction to the Theory of Complex Systems} \citep{thurner2018introduction}, {\it Agent-Based and Individual-Based Modeling: A Practical Introduction} \citep{railsback2019agent}, {\it Information Theory, Inference, and Learning Algorithms} \citep{mackay2003information}, {\it Machine Learning: A Probabilistic Perspective} \citep{murphy2022probabilistic} and {\it Ecological Networks: Linking Structure to Dynamics in Food Webs} \citep{pascual2006ecological}. Figure \ref{fig:sources_on_complex_systems} shows these sources with a small summary of their content. These references are all well-known and all of them are wonderful. However, it is worth mentioning that the field of complex systems is full of great references that we would like to consider here. {\it Complex Adaptive Systems: An Introduction to Computational Models of Social Life} \citep{miller2009complex} is one of these books. This book is a helpful guide to help us to understand how to make sense of the world with agent-based models. Another book is {\it Complexity and criticality} \citep{christensen2005complexity}. This book is a very good introduction to the theme of complexity and criticality, where a central theme of the book is scale invariance.

\begin{figure}[H] \centering \fbox{\begin{minipage}{36em}
\small
{\bf 
What are some recommended books to cover the topics of the complex systems course for PhD students?
} 

\vspace{5mm}

 {\bf ChatGPT}: There are several excellent books on complex systems that cover the topics recommended for a PhD-level course. Here are some recommended books:

"Networks, Crowds, and Markets: Reasoning About a Highly Connected World" by David Easley and Jon Kleinberg. This book provides a comprehensive introduction to network science and its applications in complex systems.

"Nonlinear Dynamics and Chaos: With Applications to Physics, Biology, Chemistry, and Engineering" by Steven H. Strogatz. This book provides a comprehensive introduction to nonlinear dynamics and chaos theory and their applications in various fields.

"Introduction to the Theory of Complex Systems" by Stefan Thurner, Rudolf Hanel, and Peter Klimek. This book provides an advanced introduction to complex systems theory, including topics such as self-organization, criticality, and complex networks.

"Agent-Based and Individual-Based Modeling: A Practical Introduction" by Steven F. Railsback and Volker Grimm. This book provides an in-depth introduction to agent-based modeling and its applications in complex systems.

"Information Theory, Inference, and Learning Algorithms" by David J. C. MacKay. This book provides an advanced introduction to information theory and its applications in complex systems.

"Machine Learning: A Probabilistic Perspective" by Kevin P. Murphy. This book provides an in-depth introduction to machine learning and its applications in complex systems analysis.

"Ecological Networks: Linking Structure to Dynamics in Food Webs" by Mercedes Pascual and Jennifer A. Dunne. This book provides an in-depth introduction to complex ecological networks and their dynamics.

These books provide a broad range of coverage of topics in complex systems, including network science, nonlinear dynamics, agent-based modeling, information theory, machine learning, and ecological networks. Depending on the course's content and the instructor's preferences, some or all of these books may be recommended for a PhD-level complex systems course.

\end{minipage}} \caption{Sources for a PhD course on complex systems.} 
\label{fig:sources_on_complex_systems} \end{figure}

\subsection{Same concept in different levels}
\label{sec:same_concept_different_level}

In this section, we explore the idea of presenting the same concept for audiences of different levels. We consider two important examples that are prominent in the field of complex systems, namely
{\it fractal} and {\it phase transition}.

\subsubsection{Fractal}
 We consider here the presentation of  the concept of {\it fractals} and {\it fractal dimension}  \cite{mandelbrot1982fractal} for different audiences. Fractals are important for complex systems modeling since many natural systems exhibit fractal-like patterns, and fractal geometry (and dimension) can be used to describe and analyze the behavior of these systems \cite{schroeder2009fractals} Furthermore, they provide a way to describe the system's behavior at different levels of detail.

We ask for an introduction of these concepts in three different levels. Figure \ref{fig:fractal_10old} shows the explanation of these concepts for a 10 year old kid, in which ChatGPT uses an example to explain them. It describes the characteristics of self-similarity of the snowflake without using this term in order to explain the concept of fractal and it compares its irregular shape with the shapes of lines, circles and squares to explain the notion of fractal dimension.

In the second case, we ask it for the same concept, but for teaching it to an undergraduate student of physics. Figure \ref{fig:fractal_under_physics} shows a more complete answer that explores explicitly the concept of self-similarity \citep{mandelbrot1967long}. It calls the attention that fractal dimensions are not integer values and it also provides the description of the {\it box counting} algorithm \citep{farmer1983dimension,feder1988fractals,falconer2004fractal}. An interesting characteristic of this answer is that ChatGPT also calls the attention for the applications of these concepts.

In the last case, we ask for teaching this concept  to a PhD student of mathematics. As we show in Figure \ref{fig:fractal_phd_math}, this answer explicitly talks about the scaling behavior of the fractal object and it provides the notion of Hausdorff-Besicovitch dimension \citep{hausdorff1918dimension,besicovitch1934sets,besicovitch1935sum}. The applications are also different from the last case. In the second case, the applications are more applied physics focused and in the third case are more general.

\begin{figure}[H] \centering \fbox{\begin{minipage}{36em}
\small
{\bf Can you help me to explain the concept of fractal and fractal dimension to a 10 year old kid?} 

\vspace{5mm}

 {\bf ChatGPT}: Sure, I'd be happy to help!
 
 Fractals are shapes that repeat themselves at different scales. That means if you zoom in on a fractal, you will see the same shape over and over again, getting smaller and smaller. They often have patterns that repeat themselves in a self-similar way.

A good example of a fractal is a snowflake. If you look closely at a snowflake, you'll notice that it has a six-pointed pattern that repeats itself as you zoom in. Each point has smaller points on it, and those points have even smaller points on them, and so on. This is what makes a snowflake a fractal.

The concept of fractal dimension is a little trickier, but I'll do my best to explain it. Fractal dimension is a way of measuring how complicated a fractal is. Unlike regular shapes like circles or squares, fractals have a dimension that is not a whole number, but a fraction.

For example, a line has a dimension of 1 because it only has one dimension, length. A square has a dimension of 2 because it has two dimensions, length and width. But a fractal like a snowflake has a dimension that is between 1 and 2, because it has some length, but also a lot of detail in the pattern that repeats itself.

So when we talk about the fractal dimension of a shape, we're talking about how much detail it has in its pattern, and how much it repeats itself at different scales.

I hope that helps! Let me know if you have any more questions.

\end{minipage}} \caption{The explanation of the concepts of fractal and fractal dimension to a 10 year old kid.} \label{fig:fractal_10old} \end{figure}

\vspace{5mm}

\begin{figure}[H] \centering \fbox{\begin{minipage}{36em}
\small
{\bf How about an undergraduate student of physics? How would you explain this concept?}

\vspace{5mm}

 {\bf ChatGPT}: For an undergraduate physics student, I can provide a more technical explanation of fractals and fractal dimension.

Fractals are geometric shapes that exhibit self-similarity at different scales. This means that if you zoom in on a fractal, you will see similar patterns repeating themselves. For example, the Mandelbrot set is a famous fractal that looks similar at all levels of magnification.

Fractal dimension is a mathematical concept used to describe the complexity of a fractal shape. Unlike traditional dimensions like length, width, and height, fractal dimension is a non-integer value that describes how much space a fractal shape occupies. This value is typically between 1 and 2, but can be greater than 2 in certain cases.

To calculate the fractal dimension of a shape, mathematicians use a process called box counting. They divide the shape into smaller and smaller boxes, and count the number of boxes that contain a part of the shape. As the size of the boxes gets smaller, the number of boxes needed to cover the shape increases. The fractal dimension is then calculated by taking the logarithm of the number of boxes needed, divided by the logarithm of the size of the boxes.

Fractal dimension is useful in physics because it can be used to describe the behavior of complex systems. For example, the fractal dimension of a coastline can give us insight into how the coast erodes over time, or the fractal dimension of a fracturing rock can give us information about its mechanical properties.

I hope this explanation helps! Let me know if you have any further questions.

\end{minipage}} \caption{The explanation of the concepts of fractal and fractal dimension to an undergraduate student of physics.} \label{fig:fractal_under_physics} \end{figure}

\begin{figure}[H] \centering \fbox{\begin{minipage}{36em}
\small
{\bf How about a phd student of mathematics? Can I give more details?
}

\vspace{5mm}

 {\bf ChatGPT}: Yes, I can certainly provide more details for a PhD student of mathematics!

Fractals are objects that have a property called self-similarity, which means that they look similar at different scales. This self-similarity can be exact or approximate, and can occur in any dimension. Some well-known examples of fractals include the Koch snowflake, the Sierpinski triangle, and the Mandelbrot set.

Fractal dimension is a way of measuring the scaling behavior of a fractal object. Unlike traditional dimensions, which are integers that describe the number of independent parameters required to specify a point in space, fractal dimension is a real number that describes the degree of irregularity or self-similarity of an object.

There are different ways to define fractal dimension, but one common method is through the Hausdorff-Besicovitch dimension. This involves covering the fractal object with small balls of radius r, and then counting the number of balls needed to cover the object. The fractal dimension is then defined as the limit of log(N(r))/log(1/r) as r approaches zero, where N(r) is the number of balls needed to cover the object of radius r.

Another way to define fractal dimension is through the concept of box-counting dimension, which is similar to the method described in the undergraduate explanation. In this method, the fractal object is covered with boxes of size r, and the number of boxes needed to cover the object is counted. The fractal dimension is then defined as the limit of log(N(r))/log(1/r) as r approaches zero.

Fractal dimension is an important concept in mathematics because it allows us to measure the complexity of fractal objects in a precise and rigorous way. It also has applications in many fields, including geometry, topology, physics, and computer science.

I hope this explanation is helpful for a PhD student of mathematics! Let me know if you have any further questions or if you would like more details on any specific aspect of the concept.

\end{minipage}} \caption{The explanation of the concepts of fractal and fractal dimension to a PhD student of mathematics.} \label{fig:fractal_phd_math} \end{figure}

\subsubsection{Phase transition}



As an another example, we can consider distinct explanations, in distinct levels, of the concept of phase transition \cite{stanley1987introduction}. Phase transitions were originally discussed in physics, in the context of liquids and magnetic systems \cite{stanley1987introduction}. However, such phenomenon was after observed in a variety of natural and artificial systems, like social, economic and biological systems. Indeed, in such distinct complex systems phase transitions are ubiquitous \cite{stauffer2006biology}.

We ask ChatGPT to introduce the main idea of phase transitions to a 10 year old kid. Figure \ref{fig:phase_transitions_10old} shows the explanation it produced. 
It uses as example the phase transitions that occur in water (liquid/solid and liquid/gas) to introduce the concept of form change. In other words, it discusses what happens with a traditional substance, water, when we heat up or cold it down.

After that, we ask how ChatGPT can explain the same concept to an undergraduate student of physics. Figure \ref{fig:phase_transitions_under_physics} shows the answer, that explores the concepts of temperature and pressure to introduce the idea of phase transition points \cite{Domb:1987eg,gitterman2014phase,yeomans1992statistical}. It also develops more details of phase transitions, like the standard classification of them in continuous or first-order transitions \cite{Jaeger1998-JAETEC}, and explain the differences between them. Finally, it calls our attention to the fact that phase transitions are very important in several areas of physics.

Finally, our last question about the subject of phase transitions to ChatGPT is how to teach this concept to a PhD student of physics. As discussed in figure \ref{fig:phase_transitions_phd_physics}, ChatGPT improves the previous answer adding the concept of criticality, which implies in the occurrence of power laws, scale invariance and critical exponents. These last concepts are indeed very important in phase transitions \cite{cardy_1996} and are useful to model a collection of applications \citep{sole2011phase,gitterman2013phase}.

\begin{figure}[H] \centering \fbox{\begin{minipage}{36em}
\small
{\bf } Can you help me to explain the concept of phase transitions to a 10 year old kid?

\vspace{5mm}

 {\bf ChatGPT}: Sure, I can help you explain the concept of phase transitions to a 10-year-old kid!

Phase transitions happen when a substance (like water) changes from one form to another form. For example, when ice melts and becomes liquid water, that's a phase transition.

When a substance changes from one form to another form, it goes through a phase transition. Imagine you have a bucket of ice. The ice is solid and very cold. But if you leave the bucket in the sun, the ice will start to melt and become liquid water. As the ice melts, it goes through a phase transition and becomes liquid.

Another example is when you boil water. When you heat up water in a pot, it starts to boil and turns into steam. This is also a phase transition.

Phase transitions can happen when you heat up a substance, cool it down, or put pressure on it. And depending on the substance and the conditions, it can go through different types of phase transitions. But the important thing to remember is that it's all about changing from one form to another form!

\end{minipage}} \caption{The explanation of the concept of phase transitions to a 10 year old kid.} \label{fig:phase_transitions_10old} \end{figure}

\vspace{5mm}

\begin{figure}[H] \centering \fbox{\begin{minipage}{36em}
\small
{\bf How about an undergraduate student of physics? How would you explain this concept?}

\vspace{5mm}

 {\bf ChatGPT}: For an undergraduate student of physics, I would explain phase transitions as follows:

Phase transitions are changes in the physical properties of a substance when it undergoes a change in state, such as from a solid to a liquid, or from a liquid to a gas. These changes occur at specific temperatures or pressures known as phase transition points.

At a phase transition point, the substance undergoes a change in its internal structure or order. For example, when water freezes and becomes ice, its molecules become more ordered and form a crystalline structure. Conversely, when ice melts and becomes liquid water, the molecules become more disordered and move more freely.

There are several types of phase transitions, including first-order transitions and continuous transitions. First-order transitions involve a change in the order of the substance, such as from a liquid to a solid, and involve a change in the enthalpy (heat content) of the substance. Continuous transitions, on the other hand, involve a change in a physical property of the substance, such as its electrical conductivity or magnetic susceptibility, and do not involve a change in enthalpy.

Phase transitions are important in many areas of physics, including materials science, condensed matter physics, and statistical mechanics. Understanding the properties and behavior of phase transitions is essential for predicting the behavior of materials under different conditions, and for developing new materials with unique properties.

\end{minipage}} \caption{The explanation of the concept of phase transition to an undergraduate student of physics.} \label{fig:phase_transitions_under_physics} \end{figure}

\begin{figure}[H] \centering \fbox{\begin{minipage}{36em}
\small
{\bf How about a phd student of physics? Can I give more details?
}

\vspace{5mm}

 {\bf ChatGPT}: Sure, I can provide more details for a PhD student of physics.

In physics, phase transitions refer to the changes in the macroscopic properties of a system when it undergoes a change in its thermodynamic state. These changes can occur as a result of temperature, pressure, or other external factors. At the phase transition point, there is a sudden change in the behavior of the system, and the properties of the system can be described by different mathematical models.

Phase transitions can be classified according to their order. First-order phase transitions involve a change in the order of the system, such as from a solid to a liquid, and are characterized by a latent heat of transition. Continuous phase transitions, on the other hand, involve a change in a physical property of the system, such as its electrical conductivity or magnetic susceptibility, and are characterized by critical exponents.

Phase transitions are of great importance in condensed matter physics, statistical physics, and materials science. They are used to understand the properties and behavior of many systems, such as magnets, superconductors, and liquid crystals. They also have practical applications in technology, such as in the design of computer chips and the development of new materials.

One of the key concepts in the study of phase transitions is criticality. Criticality refers to the behavior of a system at the critical point, where there is a sudden change in the properties of the system. At the critical point, the system exhibits scale invariance, which means that its properties are independent of the length scale. This leads to the appearance of power-law behavior, which is described by critical exponents.

In summary, phase transitions are an important concept in physics that describe the changes in the macroscopic properties of a system when it undergoes a change in its thermodynamic state. They have many applications in condensed matter physics, statistical physics, and materials science, and are characterized by criticality and critical exponents.

\end{minipage}} \caption{The explanation of the concepts of phase transition to a PhD student of physics.} \label{fig:phase_transitions_phd_physics} \end{figure}

\subsection{Connection between different concepts}
\label{sec:teaching_connection}

One important ingredient of teaching and learning is the ability to make connections between different fields. In this section, we investigate possible connections between different concepts using ChatGPT as a proxy for the complex systems and physics community. 
We discuss six examples: (1) Connection between chaos and fractals; (2) Connection between fractals and complex networks; (3) Connection between chaos and complex networks; (4) Connection between times series and complex networks; (5) Connection between chaos and machine learning.
(6) Connection between complex networks and machine learning.

\subsubsection{Connection between chaos and fractals}

In this subsection, we look into possible connections between the field of chaos \citep{lorenz1963deterministic,strogatz2018nonlinear} and fractals \citep{mandelbrot1982fractal,feder1988fractals,falconer2004fractal}. Figure \ref{fig:chaos_fractals} calls our attention to an important connection between these areas. The connection is by means of the concept of strange attractors\footnote{This term was used in the first time by \citep{ruelle1971nature}.} \citep{sprott1993strange}. An attractor is an invariant\footnote{The property of invariance means that the orbits starting in the attractor does not escape from it.}, closed\footnote{A set that contains all limit points.} and indecomposable\footnote{The set cannot be separated into smaller pieces.} set of points, which attracts all orbits of a dynamical system that start in some neighborhood (the domain of attraction) \citep{grebogi1987,HOLMES1990}. It is said to be strange if it has a non-integer fractal dimension and several researchers devised method to calculate an approximate value of this fractal dimension \citep{FROEHLING1981}.  One interesting point is that there actually needs an infinite number of dimensions to entirely characterize it \citep{HENTSCHEL1983}. 

\begin{figure}[H] \centering \fbox{\begin{minipage}{36em}
\small
{\bf Are there important connections between the fields of chaos theory and fractals?
} 

\vspace{5mm}

 {\bf ChatGPT}: Yes, there are important connections between the fields of chaos theory and fractals.

Fractals are mathematical objects that exhibit self-similarity at different scales. They are generated by iterative processes and often have a highly irregular, fragmented appearance. Chaos theory, on the other hand, studies the behavior of dynamical systems that are highly sensitive to initial conditions. These systems exhibit unpredictable and seemingly random behavior, even though they are deterministic and follow well-defined rules.

One of the important connections between chaos theory and fractals is the concept of "strange attractors". In chaotic systems, the behavior of the system may converge to a stable set of points, known as an attractor. However, if the system is highly sensitive to initial conditions, the attractor may take on a highly irregular, fractal shape. These strange attractors can be visualized using fractal geometry and are a hallmark of chaotic systems.

Fractal geometry is also used to analyze the properties of chaotic systems, such as their dimensionality. Fractal dimension is a measure of the complexity of a geometric object, and it can be used to characterize the irregular shapes of strange attractors and other features of chaotic systems.

In summary, the fields of chaos theory and fractals are closely related and share many important concepts and techniques. The study of chaotic systems often involves the use of fractal geometry to analyze and understand their behavior, and the irregular shapes of strange attractors are a common example of the connection between these fields.

\end{minipage}} \caption{Connection between chaos and fractals.} \label{fig:chaos_fractals} \end{figure}

\subsubsection{Connection between fractals and complex networks}

In this subsection, we explore possible connections between the field of fractals \citep{mandelbrot1982fractal,feder1988fractals,falconer2004fractal} and complex networks \citep{albert2002statistical,newman2006structure,cohen2010complex,estrada2012structure,posfai2016network}. Besides the fact that we can apply both fractals and complex networks to model complex systems, Figure \ref{fig:chaos_fractals} presents the use of fractal dimensions to characterize the self-similar behavior of complex networks. We may describe this connection from two different strands of the literature. The first strand refers to the literature that characterize ``complex" networks\footnote{We put complex between quotes because at that time the term ``complex network" was not actually coined yet.} embedded in Euclidean spaces, such as river networks, using fractal dimensions \citep{Tarboton1988}. An interesting point is that we may justify these scaling properties by the theory that river networks have structures that spend minimal energy in draining their watersheds \citep{Rinaldo1993}. The other strand uses fractal dimension to characterize the self-similar structures in a large number of complex networks \citep{song2005self,WEN2021}. A compelling property of fractal dimension associated with complex networks are that they can be used to characterize several other attributes of the networks such as vulnerability \citep{gou2016topological} and robustness \citep{wu2019correlation}.  

\begin{figure}[H] \centering \fbox{\begin{minipage}{36em}
\small
{\bf Are there important connections between the fields of fractals and complex networks?
} 

\vspace{5mm}

 {\bf ChatGPT}: Yes, there are important connections between the fields of fractals and complex networks. Fractals and complex networks both deal with the study of complex structures and patterns in nature and in human-made systems, and they can be used to model and understand a wide range of phenomena.

Fractals are geometric shapes that exhibit self-similarity, meaning that they look similar at different scales. They are characterized by their fractal dimension, which describes how the shape fills up space. Fractals are found in many natural systems, such as coastlines, clouds, and snowflakes, and they can also be used to model phenomena like turbulence and the behavior of financial markets.

Complex networks, on the other hand, are systems made up of many interacting elements, such as social networks, the internet, and biological systems. They are characterized by their topology, which describes how the elements are connected to each other. Complex networks are studied using tools from graph theory, and they can be used to model and understand phenomena like the spread of diseases, the behavior of financial markets, and the structure of the brain.

There are several ways in which fractals and complex networks are connected. For example:

    Fractal analysis can be used to study the structure of complex networks. By measuring the fractal dimension of different parts of a network, researchers can gain insights into its topology and how it evolves over time.
    Complex networks can be modeled using fractals. For example, researchers have used fractal geometry to model the structure of river networks and the internet.
    Fractal and complex network models can be combined to study complex systems. For example, researchers have used fractal networks to model the spread of diseases and the behavior of financial markets.

Overall, the connections between fractals and complex networks are an exciting area of research that has the potential to yield new insights into the structure and behavior of complex systems in nature and in human-made systems.

\end{minipage}} \caption{Connection between fractals and complex networks.} \label{fig:fractals_complex_networks} \end{figure}

\subsubsection{Connection between chaos and complex networks}
\label{sec:teaching_chaos_complex_networks}

In this subsection, we make connections between the fields of chaos \citep{lorenz1963deterministic,strogatz2018nonlinear} and complex networks \citep{albert2002statistical,newman2006structure,cohen2010complex,estrada2012structure,posfai2016network}. Besides the joint usage of chaos theory and complex networks to the study of pattern formation, Figure \ref{fig:chaos_complex_networks} points that chaos theory can be used to study important dynamic properties of complex, such as the emergence of synchronization  and stability \citep{li2006,wu2007synchronization,arenas2008synchronization,chen2008chaos,barrat2008dynamical}.

However, there are other  interesting attempts that connect these two fields. The main idea is to build a complex network from a chaotic dynamical system. Different methods have been used.  We may for instance represent each cycle by a node of the network \citep{zhang2006complex} or we may represent regions of space by nodes  \citep{borges2007mapping}. The advantage of modeling chaotic systems using complex networks is that we may apply the tools developed to study complex networks \citep{costa2007characterization} to understand chaotic systems.

\begin{figure}[H] \centering \fbox{\begin{minipage}{36em}
\small
{\bf Are there important connections between the fields of chaos theory and complex networks?
} 

\vspace{5mm}

 {\bf ChatGPT}: Yes, there are important connections between the fields of chaos theory and complex networks.

Chaos theory is a branch of mathematics that deals with nonlinear dynamical systems, which exhibit sensitive dependence on initial conditions, meaning that small differences in initial conditions can lead to vastly different outcomes over time. Complex networks, on the other hand, refer to systems consisting of many interacting elements, such as social networks, biological networks, and communication networks.

The study of complex networks has revealed that many real-world networks exhibit a small-world property, meaning that they have both high local clustering and short path lengths between nodes. This property is often attributed to the presence of hubs, highly connected nodes that act as major conduits for information flow within the network.

Chaos theory has been used to study the dynamics of complex networks, including the emergence of synchronization and the stability of network dynamics. One important concept in this area is the synchronization of chaotic systems, which occurs when two or more chaotic systems with similar dynamics become coordinated. This phenomenon has been observed in a variety of natural and engineered systems, including neural networks, power grids, and chemical oscillators.

Another area of research that has benefited from the connection between chaos theory and complex networks is the study of pattern formation. Many complex systems exhibit patterns that emerge from the interaction of their constituent elements, and the dynamics of these patterns can often be described using nonlinear differential equations. Chaotic systems can exhibit a wide range of patterns, including spatiotemporal chaos and fractal structures, and understanding the underlying mechanisms behind these patterns can provide insights into the behavior of complex systems.

Overall, the connections between chaos theory and complex networks provide a rich framework for studying the behavior of complex systems, and have led to numerous insights into the dynamics of natural and engineered systems.

\end{minipage}} \caption{Connection between chaos and complex networks.} \label{fig:chaos_complex_networks} \end{figure}

\subsubsection{Connection between time series and complex networks}
In this subsection, we investigate possible connections between the field of time series \citep{kantz2004nonlinear,brockwell2009time,hamilton2020time} and complex networks \citep{albert2002statistical,newman2006structure,cohen2010complex,estrada2012structure,posfai2016network}. Figure \ref{fig:time_series_complex_networks} points out that we may use complex networks to analyze time series data using tools such as the visibility graph \citep{lacasa2008time} and recurrence network methods \citep{donner2010recurrence}. As we mentioned in Section \ref{sec:teaching_chaos_complex_networks}, the advantage of these pseudo-created networks is to use techniques of the field of complex networks to model time series.  On the other hand, as Figure \ref{fig:time_series_complex_networks} also calls our attention, we may naturally use times series tools for analyzing the evolution of complex networks \citep{masuda2016guide}.

\begin{figure}[H] \centering \fbox{\begin{minipage}{36em}
\small
{\bf Are there important connections between the fields of complex networks and time series analysis?
} 

\vspace{5mm}

 {\bf ChatGPT}: 
Yes, there are important connections between the field of complex networks and time series analysis. In fact, these two fields are closely related and have been the subject of much research in recent years.

Complex networks represent systems consisting of interconnected entities, such as social networks, transportation networks, and biological networks. Time series analysis, on the other hand, deals with the study of data collected over time, such as stock prices, weather patterns, and physiological signals.

One of the main connections between these two fields is that complex networks can be used to model and analyze time series data. In particular, researchers have developed a variety of techniques for representing time series data as networks, such as the visibility graph and recurrence network methods. These techniques involve mapping the time series data onto a network, where the nodes represent the individual data points and the edges represent relationships between them.

Once the time series data has been represented as a network, researchers can use techniques from network science to analyze the data. For example, they can use measures such as degree centrality, clustering coefficient, and community structure to identify important features of the data and gain insights into its underlying dynamics.

Conversely, time series analysis can be used to study the dynamics of complex networks. In particular, researchers have used time series analysis techniques to analyze the temporal evolution of networks, such as how the connectivity between nodes changes over time.

Overall, the connections between complex networks and time series analysis are important and have led to many new insights and techniques for analyzing and understanding complex systems.

\end{minipage}} \caption{Connection between time series and complex networks.} \label{fig:time_series_complex_networks} \end{figure}

\subsubsection{Connections between chaos and machine learning}
\label{sec:chaos_machine_learning}

In this subsection, we explore possible connections between the field of chaos \citep{lorenz1963deterministic,strogatz2018nonlinear} and machine learning \citep{murphy2022probabilistic,bishop2006pattern,abu2012learning,goodfellow2016deep}. Figure \ref{fig:chaos_machine_learning} points out that we may use machine learning to model and forecast chaotic behavior \citep{Pathak2018model,Penkovsky2019coupled,amil2019machine}. However, it is not available in ChatGPT answer that chaos theory can help us to understand the dynamical behavior of recurrent neural networks behaviors \citep{doya1993bifurcations,Bengio1993problem,pmlr-v28-pascanu13}.

\begin{figure}[H] \centering \fbox{\begin{minipage}{36em}
\small
{\bf Are there important connections between the fields of chaos and machine learning?
} 

\vspace{5mm}

 {\bf ChatGPT}: Yes, there are important connections between the fields of chaos and machine learning.

Chaos theory is concerned with the study of complex, nonlinear systems that exhibit sensitive dependence on initial conditions. These systems are characterized by their unpredictable behavior over time, making it difficult to make accurate predictions about their future behavior.

Machine learning, on the other hand, is concerned with developing algorithms and models that enable computers to learn from data and make predictions or decisions based on that data. Machine learning algorithms are used in a wide range of applications, including image recognition, natural language processing, and predictive modeling.

One of the key connections between chaos and machine learning is that both fields deal with complex systems that exhibit nonlinear behavior. Machine learning algorithms can be used to model and analyze chaotic systems, enabling researchers to better understand the dynamics of these systems and make more accurate predictions about their behavior.

Another connection between the two fields is the use of neural networks. Neural networks are a class of machine learning algorithms that are loosely inspired by the structure and function of the human brain. They are capable of learning complex, nonlinear relationships between input and output data and can be used to model chaotic systems.

In summary, the connections between the fields of chaos and machine learning are significant and continue to be an area of active research. By leveraging the tools and techniques from both fields, researchers can gain new insights into complex systems and develop more accurate predictive models.

\end{minipage}} \caption{Connection between chaos and machine learning.} \label{fig:chaos_machine_learning} \end{figure}

\subsubsection{Connections between complex networks and machine learning}

In this subsection, we look into the connections between complex networks \citep{albert2002statistical,newman2006structure,cohen2010complex,estrada2012structure,posfai2016network} and machine learning \citep{murphy2022probabilistic,bishop2006pattern,abu2012learning,goodfellow2016deep}. As we may find in Figure \ref{fig:complex_networks_machine_learning}, we may use machine learning algorithms to  identify structures in a large network such as communities and important nodes.  For instance, conventional clustering algorithms may be used to identify communities \citep{FORTUNATO2010} and reinforcement learning \citep{cajueiro2009optimal,cajueiro2010optimal,kong2022}  may be used to identify important nodes. On the other hand, supervised machine learning methods that depend on attributes that come from the given network may be used for tasks as node classification  \citep{rezaei2023vital} or  link prediction \citep{LU2011,Grover2016}. There is today many machine learning methods that can be addapted to solve  diverse complex networks problems \citep{silva2016machine}. 

\begin{figure}[H] \centering \fbox{\begin{minipage}{36em}
\small
{\bf Are there important connections between the fields of complex networks and machine learning?
} 

\vspace{5mm}

 {\bf ChatGPT}: Yes, there are important connections between the fields of complex networks and machine learning.

Complex networks are a branch of network science that studies the behavior of complex systems represented as graphs, where nodes represent entities and edges represent their relationships or interactions. Machine learning, on the other hand, is a field of computer science that focuses on the development of algorithms and models that enable computers to learn from data and make predictions or decisions without being explicitly programmed.

Complex networks and machine learning are closely related because many real-world systems can be represented as complex networks, and machine learning techniques can be used to analyze and model the behavior of these networks. For example, machine learning algorithms can be used to identify patterns and structures in large-scale network data, such as community structures, important nodes or links, and network dynamics.

On the other hand, complex network analysis can also provide useful insights and features for machine learning tasks. For example, network-based features can be used as input to machine learning algorithms for prediction tasks, such as link prediction, node classification, and community detection. Additionally, network analysis techniques can be used to preprocess data for machine learning algorithms, such as data reduction or feature extraction.

In summary, the fields of complex networks and machine learning are highly interrelated, and both have much to offer in terms of insights, models, and algorithms for understanding and analyzing complex systems.

\end{minipage}} \caption{Connection between complex networks and machine learning.} \label{fig:complex_networks_machine_learning} \end{figure}

\subsection{Learning activities}
\label{sec:teaching_learning_activities}

Learning activities, such as hands-on and experiential learning experiences, encourage active learning by engaging students. In this section, we explore the possibility of ChatGPT suggesting an interesting learning activity to help individuals learn an important concept in the complex systems field. Thus, we ask ChatGPT a learning activity to teach children about consensus in opinion dynamics \citep{xia2011opinion} and it suggests a game that the objective is to reach a consensus. 

\begin{figure}[H] \centering \fbox{\begin{minipage}{36em}
\small
{\bf What is a learning activity that can be used to teach children about consensus in opinion models?
} 

\vspace{5mm}

 {\bf ChatGPT}: 
A great learning activity to teach children about consensus in opinion models is to use a group decision-making game. Here's an example of a simple game you can use:

Divide the children into small groups of 3-4.
Give each group a problem to solve or a decision to make. For example, you could ask them to choose a color for a new classroom poster or decide on a game to play during recess.
Ask each child in the group to write down their individual choice on a piece of paper.
Have the children share their choices with each other, and encourage them to discuss why they made their choice.
Ask the group to work together to come up with a single choice that everyone can agree on. Encourage them to listen to each other's opinions and try to find common ground.
Once the group has reached a consensus, ask them to share their final decision with the class.
This activity will help children understand the importance of listening to others and working together to reach a common goal. It will also teach them about the concept of consensus and how it can be used to make decisions as a group.

\end{minipage}} \caption{Learning activity to understand the concept of consensus.} \label{fig:learning_actitivity_consensus} \end{figure}

\subsection{Practical laboratory experiments}
\label{sec:teaching_lab}
In addition to lectures and other learning activities, teachers may schedule laboratory experiments as practical activities to help students learn new concepts in the complex systems field. 

First, we want to know how to design practical laboratory experiments in the field of complex systems and, then, we ask for the suggestion of such experiments. Figure \ref{fig:lab_experiments} presents the basic steps for designing practical laboratory experiments. It considers the identification of the concepts we want to demonstrate, the correct choice of equipment and materials, a previous test of the experiment, safety issues, interactiveness, data visualization and simplicity.

Second, we want to know what specific experiments we should consider. Figure \ref{fig:lab_experiments_suggestion} shows some interesting suggestions, such as the {\it bouncing  ball experiment} \citep{brancazio1981physics,cross1999bounce} to demonstrate sensitivity to initial conditions ubiquitous in chaotic systems, {\it  a traffic flow simulation} \citep{nagel1995emergent,helbing2001self,Treiber2012Traffic} to show the emergent behavior of complex systems, a {\it network analysis} \citep{costa2007characterization,caldarelli2016data}, such as social \citep{barabasi1999emergence,vega2007complex} or transportation networks \citep{latora2002boston,cajueiro2009optimal}, to investigate the structure of a complex network, a {\it pattern formation in a chemical reaction} to reveal the formation of complex patterns from simple chemical reactions \citep{kuramoto1976pattern,cross1993pattern}, and  a {\it swarm robotics} \citep{bonabeau2000swarm,Spezzano2019swarm} experiment where robots work together to realize a task to show how complex systems are formed from simple components.

\begin{figure}[H] \centering \fbox{\begin{minipage}{36em}
\small
{\bf How can I design practical laboratory experiments in the field of complex systems that effectively demonstrate key concepts and principles?
} 

\vspace{5mm}

 {\bf ChatGPT}: Designing practical laboratory experiments in the field of complex systems requires careful consideration of a variety of factors, such as the specific principles and concepts that you want to demonstrate, the equipment and materials available, and the level of experience of your audience. Here are some general tips for designing effective laboratory experiments in this field:

Identify the key concepts and principles you want to demonstrate: Start by clearly defining the concepts and principles you want to teach, and make sure they are appropriate for the level of your audience. Complex systems can encompass a wide range of topics, such as chaos theory, network dynamics, or emergence, so it's important to choose a few key concepts to focus on.

Choose appropriate equipment and materials: Based on the concepts you want to teach, select the appropriate equipment and materials for your experiment. You may need specialized software, sensors, or data collection tools to conduct your experiment effectively.

Test your experiment beforehand: Before conducting the experiment in front of your audience, it's essential to test it beforehand to ensure that it works as intended. This will also give you an opportunity to identify any potential problems and make adjustments as needed.

Ensure safety: When working with laboratory equipment, safety should always be a top priority. Make sure you and your audience are familiar with any safety protocols or precautions, and take steps to minimize risks.

Make it interactive: Complex systems can be abstract and difficult to visualize, so it's important to make your experiment as interactive as possible. Encourage your audience to ask questions and participate in the experiment, and provide opportunities for them to manipulate variables and see how the system responds.

Use data visualization: Complex systems often generate large amounts of data, so it's important to use data visualization tools to help your audience understand the patterns and trends that emerge. This can include graphs, charts, or other visual representations of the data.

Keep it simple: While complex systems can be fascinating, it's important to keep your experiment simple and focused on a few key concepts. Too many variables or too much complexity can be overwhelming for your audience and may make it difficult for them to understand the principles you're trying to teach.

By following these tips, you can design practical laboratory experiments in the field of complex systems that effectively demonstrate key concepts and principles. Remember to always be mindful of safety and to tailor your experiments to the level of your audience.

\end{minipage}} \caption{Design of laboratory experiments.} \label{fig:lab_experiments} \end{figure}

\begin{figure}[H] \centering \fbox{\begin{minipage}{36em}
\small
{\bf Can you suggest some practical experiments?

}

\vspace{5mm}

 {\bf ChatGPT}: Sure, here are a few examples of practical experiments that you could conduct in the field of complex systems:

Bouncing ball experiment: This experiment is a classic example of chaos theory and demonstrates how small changes in initial conditions can lead to vastly different outcomes over time. In this experiment, you drop a ball onto a surface and measure the height of each bounce. By making small changes to the initial conditions (such as the height of the drop or the angle of the surface), you can observe how the behavior of the system changes over time.

Traffic flow simulation: This experiment demonstrates how simple rules can lead to emergent behavior in complex systems. In this experiment, you create a simulation of traffic flow using a simple set of rules (such as maintaining a safe following distance and adjusting speed based on traffic density). By changing the parameters of the simulation, you can observe how the behavior of the system changes over time.

Network analysis: This experiment involves analyzing the structure and behavior of networks (such as social networks or transportation networks) to understand how information or resources flow through them. You could use network analysis software to analyze the structure of a network and identify key nodes or pathways.

Pattern formation in chemical reactions: This experiment demonstrates how complex patterns can emerge from simple chemical reactions. In this experiment, you mix two or more chemicals and observe how the reaction progresses over time. By changing the concentrations of the chemicals or the temperature of the reaction, you can observe how different patterns emerge.

Swarm robotics: This experiment involves programming a group of robots to work together to accomplish a task (such as moving objects or exploring an environment). By observing how the robots interact and how their behavior changes over time, you can gain insights into how complex systems can be built from simple components.

These are just a few examples of practical experiments in the field of complex systems. Depending on your specific interests and the resources available to you, there are many other experiments you could design and conduct.

\end{minipage}} \caption{Suggestions of laboratory experiments.} \label{fig:lab_experiments_suggestion} \end{figure}

\subsection{Coding}
\label{sec:teaching_code}

An important dimension of teaching and learning complex systems theory is coding. Figure \ref{fig:lis_python} shows the Python libraries that may be useful for complex systems' researchers. It points to {\it NumPy} \citep{harris2020array} for numerical tools, {\it SciPy} \citep{2020SciPy-NMeth} for statistical tools, {\it NetworkX}  \citep{hagberg2008exploring} for complex networks modelling, {\it Matplotlib} \citep{Hunter:2007} for plotting, {\it Pandas} \citep{mckinney-proc-scipy-2010} for struturing and analysing data, {\it SymPy} \citep{10.7717/peerj-cs.103} for symbolic tools, {\it PyCX} \citep{sayama2013pycx} for complex systems simulations and {\it Scikit-learn} \citep{scikit-learn} for machine learning models. Although it provides a great portfolio of libraries, we miss the {\it Mesa} library \citep{python-mesa-2020}, that is a great library for agent-based models simulation. Thus, we ask about the Mesa library. Figure \ref{fig:python_library_mesa}  shows the relevance of this library for complex systems researchers. Finally, we ask ChatGPT to provide an example of use of the NetworkX library. Figures \ref{fig:networkx1}, \ref{fig:networkx2} and \ref{fig:networkx3} provide a simples example of building, characterizing and plotting  a social network for self-learning\footnote{It is worth mentioning that the code it provides has a minor execution error, since the social network it builds is not fully connected and we need this property to evaluate the average shortest path. Basically, we have to include the pair {\it ('Frank','Alice')} in the list {\it friendships} before running, but we note that the syntatix is perfect.}.

\begin{figure}[H] \centering \fbox{\begin{minipage}{36em}
\small
{\bf Which python libraries may be useful for researchers of the field of complex systems?

} 

\vspace{5mm}

 {\bf ChatGPT}: Python has a rich set of libraries that can be useful for researchers working in the field of complex systems. Here are some of the most commonly used ones:

NumPy: This library provides support for large, multi-dimensional arrays and matrices, and offers a range of mathematical functions to work with them. This is particularly useful for complex systems research where large amounts of data need to be processed and analyzed.

SciPy: This library builds on top of NumPy and provides additional functions for scientific and technical computing, such as optimization, integration, and linear algebra.

NetworkX: This library provides tools for the creation, manipulation, and study of complex networks. It offers functions for generating random graphs, computing network metrics, and visualizing network structures.

Matplotlib: This library provides a wide range of tools for creating visualizations and plots, which can be used to illustrate complex system behavior.

Pandas: This library provides tools for data manipulation and analysis. It offers data structures for efficient handling of large datasets and functions for statistical analysis.

SymPy: This library provides support for symbolic mathematics, which is particularly useful for researchers working in the field of mathematical modeling of complex systems.

PyCX: This library provides a set of tools for creating simulations of complex systems, which can help researchers to explore and analyze the behavior of such systems.

Scikit-learn: This library provides a range of tools for machine learning, which can be useful for analyzing and predicting the behavior of complex systems.

There are many more libraries available for Python, but these are some of the most commonly used ones for researchers working in the field of complex systems.

\end{minipage}} \caption{List of Python libraries useful for complex systems' researchers.} \label{fig:lis_python} \end{figure}

\begin{figure}[H] \centering \fbox{\begin{minipage}{36em}
\small
{\bf Don't you think the mesa library can be useful?

} 

\vspace{5mm}

 {\bf ChatGPT}: Yes, the Mesa library can be very useful for researchers working in the field of complex systems. Mesa is a Python library for agent-based modeling (ABM), which is a powerful technique for simulating and analyzing the behavior of complex systems that are made up of individual agents that interact with one another.

Mesa provides a framework for creating, running, and analyzing agent-based models, and includes a range of tools and utilities for building and visualizing these models. Some of the key features of Mesa include:

A simple and flexible API for defining agents and their behaviors, which allows researchers to easily implement and test new models.

Built-in tools for creating visualizations and animations of the models, which can be useful for understanding the dynamics of complex systems.

A range of utility functions for analyzing and visualizing the data generated by the models, which can help researchers to identify patterns and trends in the behavior of the agents.

Support for parallel execution, which allows the models to be run efficiently on multi-core processors and distributed computing clusters.

Some examples of applications of the Mesa library include modeling the spread of diseases, analyzing the behavior of financial markets, and simulating the interactions between animals in an ecosystem.

Overall, if you are working in the field of complex systems and are interested in agent-based modeling, the Mesa library can be a valuable tool for building and analyzing models of complex systems.

\end{minipage}} \caption{The relevance of Mesa library.} \label{fig:python_library_mesa} \end{figure}

\begin{figure}[H] \centering \fbox{\begin{minipage}{36em}
\small
{\bf Can you provide an interesting example of the use of the library NetworkX?
} 

\vspace{5mm}

 {\bf ChatGPT}: 

Sure! Here's an example of how NetworkX can be used to analyze the structure of a social network:

Suppose we have data on the friendships between a group of individuals, and we want to analyze the structure of the resulting social network. We can represent the network as a graph, where each individual is a node and each friendship is an edge. We can use NetworkX to create this graph and analyze its properties. Here's how we can do it:

First, we need to create the graph. We can use the Graph() function from NetworkX to create an empty graph, and then add nodes and edges to it based on our data. Here's some code to do this:

\end{minipage}}

\small
\begin{mdframed}[backgroundcolor=lightgray]
\begin{verbatim}
import networkx as nx

# Create an empty graph
G = nx.Graph()

# Add nodes to the graph for each individual
individuals = ['Alice', 'Bob', 'Charlie', 'David', 
               'Eve', 'Frank']
for i in individuals:
    G.add_node(i)

# Add edges to the graph for each friendship
friendships = [('Alice', 'Bob'), ('Alice', 'Charlie'), 
                ('Bob', 'Charlie'), ('David', 'Eve'), 
                ('Eve', 'Frank')]
for i, j in friendships:
    G.add_edge(i, j)

%\end{minted}
\end{verbatim}
\end{mdframed}

\caption{Example with the NetworkX library: creation of the network.} \label{fig:networkx1} \end{figure}

\begin{figure}[H] \centering \fbox{\begin{minipage}{36em}
\small
Once we have created the graph, we can analyze its properties. For example, we can compute the degree distribution of the nodes in the graph, which tells us how many edges each node has. We can also compute the average shortest path length between nodes, which tells us how connected the graph is. Here's some code to do this:

\end{minipage}} 
\begin{mdframed}[backgroundcolor=lightgray]

\small
\begin{verbatim}
# Compute the degree distribution
degrees = [d for n, d in G.degree()]
degree_counts = nx.degree_histogram(G)

# Compute the average shortest path length
avg_shortest_path = nx.average_shortest_path_length(G)

# Print the results
print('Degree distribution:', degree_counts)
print('Average shortest path length:', avg_shortest_path)

%\end{minted}
\end{verbatim}
\end{mdframed}

\caption{Example with the NetworkX library: characterization of the network.} \label{fig:networkx2} \end{figure}

\begin{figure}[H]
\end{figure}

\begin{figure}[H] \centering \fbox{\begin{minipage}{36em}
\small

Finally, we can visualize the graph using NetworkX's built-in plotting functions. We can use a spring layout to position the nodes, which will make the graph look more natural. Here's some code to do this:

\end{minipage}} 
\begin{mdframed}[backgroundcolor=lightgray]
\small
\begin{verbatim}
import matplotlib.pyplot as plt

# Compute the layout using a spring layout algorithm
pos = nx.spring_layout(G)

# Plot the graph
nx.draw_networkx_nodes(G, pos, node_color='lightblue')
nx.draw_networkx_edges(G, pos, edge_color='gray')
nx.draw_networkx_labels(G, pos)
plt.axis('off')
plt.show()

%\end{minted}     
\end{verbatim}
\end{mdframed}

 \fbox{\begin{minipage}{36em}
\small

This code will display a visualization of the social network, where each node represents an individual and each edge represents a friendship. The layout of the nodes is determined by a spring layout algorithm, which tries to position the nodes so that they are evenly spaced and not too close together. We can use this visualization to gain insights into the structure of the social network and to identify key individuals or communities within it

\end{minipage}}

\caption{Example with the NetworkX library: plot of the network.} \label{fig:networkx3} \end{figure}

\section{Research}

Researchers, specially those who are outsiders to a particular field, face regularly the challenge of keeping up with the latest developments and understanding the subject matter. 
In this section, we use ChatGPT  as a proxy for the answers that can be found within the physics and complex systems community to spot the purpose, trends, critiques and challenges of the field. 
Section \ref{sec:research_overview} presents an overview of the complex systems field identifying the purpose and trends of the field. Besides focusing only on the entire complex systems field, we also explore the subfields of econophysics, geographic complexity and complexity in biological systems. While in Section \ref{sec:research_complex_systems_critique}, we investigate the critiques to the field, in Section \ref{sec:research_challenges}, we discuss the challenges of the field.

\subsection{Overview of the field}
\label{sec:research_overview}
We split this subsection into three subsections. While Subsection \ref{sec:complex_systems_overview} explores the field of complex systems, Subsections \ref{sec:econophysics} and \ref{sec:geographic_complexity} investigate respectively the subfields of econophysics and geographic complexity.

\subsubsection{Complex systems}
\label{sec:complex_systems_overview}

We start asking ChatGPT the purpose of the field of complex systems. Figure \ref{fig:complex_systems_purpose} shows the basic relevant ingredients of the field \cite{miller2009complex,kwapien2012physical,charbonneau2017natural,thurner2018introduction}, which is very connected with the definition presented in Section \ref{sec:definition_complex_systems}. 
It also calls the attention that the field is a multidisciplinar one and it may be applied in different disciplines. It concludes that another important function of the field is to develop strategies for managing and controlling complex systems. 

In the following, we ask ChatGPT about the current trends and developments of the field. We may find in Figure \ref{fig:complex_systems_trends} the answer to this question that highlights the trends of interest in {\it network science} \citep{albert2002statistical,newman2006structure,cohen2010complex,estrada2012structure,posfai2016network}, {\it artificial intelligence and machine learning} \citep{jordan2015machine,silva2016machine,Tang2020introduction}, {\it agent-based modelling} \citep{bonabeau2002agent,helbing2012agent,de2014agent}, {\it complexity economics} \citep{stanley1996anomalous,mantegna1999introduction,Gligor2001econophysics,Schinckus2013introduction,slanina2014} and {\it interdisciplinary collaboration} \citep{Newell2001,haken2012interdisciplinary}. Note that the first three topics are topics already suggested as topics for a course on complex systems in Section \ref{sec:teaching_topic}. {\it Complexity economics} or {\it Econophysics} is a subfield of the field of complex systems and we explore it in Section \ref{sec:econophysics}. Finally, it is noteworthy that the trend of interdisciplinary collaboration in complex systems is gaining momentum, as experts from various fields recognize the importance of complex systems methods and seek to apply them in their own work. In addition, collaboration from other fields is highly encouraged, as the participation of experts from diverse fields can enhance the credibility and reputation of research in the complex systems field.

\begin{figure}[H] \centering \fbox{\begin{minipage}{36em}
\small
{\bf What is the purpose of the complex systems field?

} 

\vspace{5mm}

 {\bf ChatGPT}: The purpose of the complex systems field is to study and understand the behavior of complex systems, which are systems made up of many interconnected and interdependent components that exhibit non-linear, dynamic, and often unpredictable behavior.

The complex systems field seeks to develop theories, models, and methods to analyze and predict the behavior of complex systems, such as the economy, ecosystems, social networks, biological systems, and many others. The field draws on a wide range of disciplines, including mathematics, physics, computer science, biology, and social sciences, to develop a multidisciplinary approach to understanding complex systems.

One of the goals of the complex systems field is to identify patterns and emergent properties that arise from the interactions of the individual components of a complex system. These patterns and properties may not be apparent when looking at the individual components in isolation, but can be observed at the system-level.

Another important purpose of the complex systems field is to develop strategies for managing and controlling complex systems, especially in the face of uncertainty and change. This may involve identifying critical components or feedback loops within a system that can be targeted for intervention, or developing adaptive strategies that can respond to changes in the system over time.

\end{minipage}} \caption{The purpose of the complex systems field.} \label{fig:complex_systems_purpose} \end{figure}

\begin{figure}[H] \centering \fbox{\begin{minipage}{36em}
\small
{\bf Which are the current trends and developments in the complex systems field?

} 

\vspace{5mm}

 {\bf ChatGPT}: The complex systems field is a rapidly evolving and dynamic field, and there are many current trends and developments that are shaping the direction of research and applications in the field. Some of these trends and developments include:

Network Science: Network science is a subfield of complex systems that focuses on the study of complex networks, which are systems of interconnected entities. Network science is being applied to a wide range of fields, including social networks, transportation networks, biological networks, and many others.

Artificial Intelligence and Machine Learning: Advances in artificial intelligence and machine learning are enabling researchers to develop more sophisticated models of complex systems, and to better understand the behavior of these systems. AI and machine learning techniques are also being used to develop more effective strategies for managing and controlling complex systems.

Agent-Based Modeling: Agent-based modeling is a simulation technique that is used to study the behavior of complex systems by modeling the behavior of individual agents within the system. Agent-based modeling is being used to study a wide range of complex systems, including ecosystems, transportation systems, and social networks.

Complexity Economics: Complexity economics is a relatively new field that applies complex systems thinking to the study of economic systems. Complexity economics seeks to develop a better understanding of the behavior of economic systems, including how they evolve and how they respond to shocks and changes.

Interdisciplinary Collaboration: Collaboration between researchers from different disciplines is becoming increasingly important in the complex systems field. By bringing together experts from different fields, researchers can develop more comprehensive and integrated approaches to studying complex systems.

Overall, the complex systems field is a rapidly evolving and dynamic field that is making important contributions to our understanding of complex systems and their behavior. As new tools and techniques continue to emerge, it is likely that the field will continue to grow and develop in new and exciting ways.

\end{minipage}} \caption{Current trends and developments of the complex systems field.} \label{fig:complex_systems_trends} \end{figure}

\subsubsection{Econophysics}
\label{sec:econophysics}

In this section, we explore the field of econophysics that is a subfield of the field of complex systems. We start by asking ChatGPT about
the purpose of Econophysics. We show the answer to this question in Figure \ref{fig:econophysics_purpose}, where it asserts (with a very complete answer) that econophysics is that uses tools of complex systems to investigate economic behavior \citep{stanley1996anomalous,mantegna1999introduction,Gligor2001econophysics,Schinckus2013introduction,slanina2014,kutner2019econophysics}. 

Next, we ask ChatGPT for the possible trends in Econophysics. Figure \ref{fig:econophysics_trends} stresses the trends of interest in  
{\it network theory} \citep{Mantegna1999hierarchical,boss2004,tabak2014directed}, {\it machine learning} \citep{BASALTO2007,cajueiro2021model}, {agent-based modelling} \citep{Gatti2018,lux2021}, {\it nonlinear-dynamics} \cite{HSIEH1991,sornette2017stock} and {\it complexity economics} \citep{hidalgo2009building,hidalgo2021economic}. 

We know that the study of persistence in financial time series is largely investigated \citep{costa2003long,cajueiro2004hurst,cont2005long,morales2012dynamical,salcedo2022persistence}. Thus, we ask if this topic cannot be considered a trend anymore. In Figure \ref{fig:econophysics_persistence} mentions that although this is not a new trend anymore, it agrees that it is an active and important topic since it has implications for risk management and the behavior of financial markets. Furthermore, it presents some of the techniques used to run these investigations such as the Hurst exponent \citep{hurst1951long,feder1988fractals} and Detrended Fluctuation Analysis (DFA) \citep{moreira1994fractal,peng1994mosaic}.

In the following, we ask about a possible impact of the field of econophysics in mainstream economics. Figure \ref{fig:econophysics_impact} presents the answer to this question that defends that measuring this impact is difficult, since the field is still relatively new and its ideas are not yet widely accepted in the mainstream. Anyway, it suggests three criteria, namely the amount of citations that econophysics papers receive from mainstream economics papers, the amount of collaboration between the two fields and the extent econophysics models are adopted by mainstream economics. We have to agree that these criteria are really good suggestions. 

Thus, we ask ChatGPT the reason that the ideas of econophysics are not widely accepted in mainstream economics. Figure \ref{fig:econophysics_not_accepted} shows the answer to this question that provides four different reasons: (1) {\it Different approaches} (methods from physics versus assumptions about individual agent behavior); (2) {\it Lack of empirical validation}\footnote{This is the classical critique to agent-based models. However, it is worth mentioning that in the last few years, some researchers have proposed solutions to this issue \citep{windrum2007empirical,FRANKE2012structural,peralta2022opinion}.}; (3) {\it Resistance to change}; (4) {\it Lack of familiarity with the methods used in econophysics}. We agree with all of them. 
In particular, due to the difficulty to run large scale experiments in economics, the field is very conservative and, therefore, resistant to change. However, we want to highlight that maybe the first item is the most relevant. Although this is not explicit in the first item, the primary objective of mainstream economics is to model the choice behavior of agents and many of the econophysics models do not follow this objective \citep{Barroso_et_all2016,ADAO2022impacts}. 

In spite of the resistance of mainstream economics to the field of econophysics, it is worth mentioning that some works published in high impact economic journals could be by members of the community of econophysics.  Among them we may cite {\it Self-organized criticality and economic fluctuations} \citep{scheinkman1994self}, {\it The granular origins of aggregate fluctuations} \citep{gabaix2011granular} and {\it The dynamics of inequality} \citep{gabaix2016dynamics}.

\begin{figure}[H] \centering \fbox{\begin{minipage}{36em}
\small
{\bf What is the purpose of econophysics?
} 

\vspace{5mm}

 {\bf ChatGPT}: Econophysics is a field of study that applies concepts and tools from physics to economic systems. Its purpose is to use methods from physics, such as statistical mechanics, network theory, and complex systems theory, to model and analyze complex economic phenomena.

The goal of econophysics is to develop a deeper understanding of economic systems, including financial markets, income distributions, and macroeconomic dynamics, by studying them as complex systems. By doing so, econophysicists hope to uncover underlying patterns and structures that may be difficult to detect using traditional economic approaches.

Econophysics also aims to provide insights into the behavior of economic systems during times of crisis, such as stock market crashes or economic recessions, and to develop models that can help policymakers make more informed decisions.

\end{minipage}} \caption{The purpose of the econophysics field.} \label{fig:econophysics_purpose} \end{figure}

\begin{figure}[H] \centering \fbox{\begin{minipage}{36em}
\small
{\bf Which are the current trends and developments in econophysics?

} 

\vspace{5mm}

 {\bf ChatGPT}: There are several current trends and developments in the field of econophysics, some of which are:

Network theory: Econophysicists are increasingly using network theory to study economic systems. This involves modeling economic agents as nodes in a network, and the connections between them as edges. By analyzing the properties of these networks, researchers can gain insights into the structure of economic systems, including financial markets.

Machine learning: Machine learning techniques are being increasingly used in econophysics to analyze large and complex data sets. This includes techniques such as clustering, regression, and deep learning.

Agent-based modeling: Agent-based modeling is a technique used in econophysics to simulate the behavior of economic agents, such as investors or firms, in a complex system. This allows researchers to explore the emergent properties of the system and make predictions about its behavior.

Nonlinear dynamics: Nonlinear dynamics is a branch of mathematics that deals with the behavior of systems that are sensitive to initial conditions. Econophysicists are using this approach to study the dynamics of financial markets, including the emergence of bubbles and crashes.

Complexity economics: Complexity economics is a new approach that seeks to understand economic systems as complex adaptive systems. This involves studying the interactions between economic agents and the emergence of patterns and structures that are not easily predicted from individual behavior.

Overall, these trends and developments in econophysics are helping to provide new insights into the behavior of economic systems and inform policy decisions.

\end{minipage}} \caption{Current trends and developments of the econophysics field.} \label{fig:econophysics_trends} \end{figure}

\begin{figure}[H] \centering \fbox{\begin{minipage}{36em}
\small
{\bf How about the topic that investigates persistence in financial time series? Cannot it be considered a trend anymore?

} 

\vspace{5mm}

 {\bf ChatGPT}: The investigation of persistence in financial time series is still a relevant and active area of research in econophysics. Persistence refers to the tendency of financial variables, such as asset prices or volatility, to exhibit long-term correlations over time. This is an important phenomenon to understand because it can have implications for risk management and the behavior of financial markets.

Econophysicists have developed several techniques to analyze persistence in financial time series, such as detrended fluctuation analysis (DFA) and the Hurst exponent. These techniques can help to identify whether a time series exhibits long-term correlations and to quantify the degree of persistence.

While the investigation of persistence in financial time series is not a new trend in econophysics, it remains an important topic of study. There is ongoing research aimed at understanding the causes of persistence, its effects on financial markets, and how it can be modeled and predicted. Therefore, this topic can still be considered an active area of research in the field of econophysics.

\end{minipage}} \caption{The relevance of the topic that investigates persistence in financial time series.} \label{fig:econophysics_persistence} \end{figure}

\begin{figure}[H] \centering \fbox{\begin{minipage}{36em}
\small
{\bf Is it possible to measure the impact of econophysics in mainstream economics?

} 

\vspace{5mm}

 {\bf ChatGPT}: Measuring the impact of econophysics on mainstream economics can be challenging, as the field is still relatively new and its ideas are not yet widely accepted in the mainstream. However, there are some ways to assess its impact:

Citation analysis: One way to measure the impact of econophysics on mainstream economics is to analyze the number of citations that econophysics papers receive from mainstream economics journals. While this is not a perfect measure, it can give an indication of the level of attention and recognition that econophysics is receiving.

Collaboration and interdisciplinary research: Another way to measure the impact of econophysics is to look at the level of collaboration and interdisciplinary research between econophysicists and mainstream economists. If there are more joint research projects or conferences, it could be an indication of a growing interest in econophysics from mainstream economics.

Adoption of econophysics models: A third way to measure the impact of econophysics is to look at the extent to which econophysics models and concepts are being adopted in mainstream economic research. For example, if mainstream economists are using network theory or agent-based models in their research, it could be an indication that econophysics is having an impact.

Overall, while it may be difficult to measure the impact of econophysics on mainstream economics, there are some indications that its ideas and models are gaining attention and recognition in the field.

\end{minipage}} \caption{Impact of the field of econophysics in mainstream economics.} \label{fig:econophysics_impact} \end{figure}

\begin{figure}[H] \centering \fbox{\begin{minipage}{36em}
\small
{\bf Why the ideas of econophysics are not widely accepted in mainstream economics?
} 

\vspace{5mm}

 {\bf ChatGPT}: 
The ideas of econophysics are not widely accepted in mainstream economics for several reasons:

Different approaches: Econophysics and mainstream economics have different approaches to studying economic systems. Econophysics uses methods from physics and complex systems theory to model and analyze economic phenomena, while mainstream economics typically uses mathematical models based on assumptions about individual behavior.

Lack of empirical validation: Some mainstream economists are skeptical of econophysics because many of its models have not yet been empirically validated. While econophysics models have shown promise in explaining certain phenomena, such as stock market behavior, there is still much work to be done to validate these models and demonstrate their usefulness.

Resistance to change: Like any field, mainstream economics has its established ways of doing things, and it can be difficult for new ideas and approaches to gain acceptance. Econophysics is a relatively new field, and it may take time for its ideas to be accepted and integrated into mainstream economics.

Complexity and interdisciplinary nature: Econophysics is an interdisciplinary field that draws on ideas and methods from physics, mathematics, and computer science, among other disciplines. This can make it difficult for economists who are not familiar with these fields to understand and appreciate econophysics models and ideas.

Overall, while there are several reasons why the ideas of econophysics are not widely accepted in mainstream economics, there is growing interest in the field and the potential for it to provide new insights into economic phenomena.

\end{minipage}} \caption{The reasons that econophysics' ideas are widelly accepted in economics.} \label{fig:econophysics_not_accepted} \end{figure}

\subsubsection{Geographic complexity}
\label{sec:geographic_complexity}

In this section, we explore another subfield of the complex systems field known as {\it geographic complexity} \citep{batten2001complex,batty2008size,de2010planner,barthelemy2019statistical}. As in the case of econophysics considered in Section \ref{sec:econophysics}, we start by asking ChatGPT about the purpose of this field. Figure \ref{fig:geographic_complexity_purpose} shows the answer to this question that says it is an interdisciplinary field with the objective of investigating how physical and human systems interact with each other across various scales of space and time. There other interesting ingredients in the answer to this question, such as the types of phenomena the field intends to model and the methodological approaches used in the field. The field uses methods from  {\it geographic information science} \citep{longley1997spatial}, {\it spatial analysis} \citep{bailey1995interactive}, {\it complexity theory} \citep{white2015modeling}, and {\it simulation modeling}  to investigate phenomena such as  {\it climate change} \citep{rind1999complexity}, {\it land-use change} \citep{Crawford2005,Rindfuss2008land}, {\it natural disasters} \citep{RUS2018}, {\it economic development} \citep{Balland2017,hartmann2022economic}, and {\it urbanization} \citep{batty1976urban,makse1998modeling,BATTY1999modeling,oliveira2014large}. 

In the following, we ask about the current trends and developments of this field. Figure \ref{fig:geographic_complexity_trends} shows the answer to this question that points the following trends: {\it big data and machine learning} \citep{dobesova2020experiment,lavallin2021machine},   {\it integration of physical and social systems} \citep{o2015physicists}, {\it multi-scale analysis} \citep{bilal2021scaling,de2020socio}, {\it collaborative modeling} \citep{pumain1992geography} and {\it Open science and reproducibility} \citep{Nust2021,kedron2021}. The trend of utilizing big data and machine learning techniques is widespread due to the increased availability of large datasets and advancements in data analysis methods. The trend of open science and reproducibility is also a widespread phenomenon in empirical sciences caused by the reproducibility crisis. In addition, it fosters transparency and efficiency that saves time and resources. The trend of collaborative modeling is associated with the same reasons of the trend found in the complex systems literature. The trend of big data seems to be continuous trend in this literature that explores geographic phenomena at different  scales.

\begin{figure}[H] \centering \fbox{\begin{minipage}{36em}
\small
{\bf What is the purpose of the field known as geographic complexity?
} 

\vspace{5mm}

 {\bf ChatGPT}: Geographic complexity is an interdisciplinary field that seeks to understand the complex and dynamic relationships between physical and human systems at different spatial and temporal scales. The purpose of this field is to provide insights into the complex nature of geographical phenomena, such as climate change, land-use change, natural disasters, economic development, and urbanization, among others.

Geographic complexity draws on various theoretical and methodological approaches, including geographic information science, spatial analysis, complexity theory, and simulation modeling, to understand the interplay between natural and human systems. It also seeks to provide practical solutions to real-world problems related to spatial planning, resource management, and sustainable development.

Overall, the purpose of geographic complexity is to provide a more comprehensive and nuanced understanding of the complex spatial dynamics that shape our world, and to develop strategies that promote more sustainable and equitable outcomes for people and the planet.

\end{minipage}} \caption{The purpose of the geographic complexity field.} \label{fig:geographic_complexity_purpose} \end{figure}

\begin{figure}[H] \centering \fbox{\begin{minipage}{36em}
\small
{\bf Which are the current trends and developments in the geographic complexity field?

} 

\vspace{5mm}

 {\bf ChatGPT}: There are several current trends and developments in the geographic complexity field. Here are a few examples:

Big data and machine learning: The availability of large and complex datasets, combined with advances in machine learning algorithms, is revolutionizing the way geographers analyze and model complex spatial phenomena. This has led to the development of new tools and techniques for spatial analysis, such as deep learning, network analysis, and agent-based modeling.

Integration of physical and social systems: One of the key challenges in the geographic complexity field is to understand the interplay between physical and social systems. Recent developments in this area have focused on developing more integrated models that can capture the feedback loops and dynamics between these systems, such as land-use change and urbanization.

Multi-scale analysis: Many geographic phenomena operate at different spatial and temporal scales, and understanding how they interact requires a multi-scale approach. Advances in GIS and remote sensing technologies have made it possible to collect and analyze data at increasingly fine scales, while also providing new tools for integrating data across scales.

Collaborative modeling: Collaborative modeling involves engaging stakeholders and experts from different fields to develop models and scenarios that reflect their diverse perspectives and goals. This approach has gained popularity in recent years as a way to develop more robust and effective strategies for spatial planning and resource management.

Open science and reproducibility: The geographic complexity field has increasingly embraced open science principles, which promote the sharing of data, code, and methods to promote transparency and reproducibility in research. This has led to the development of new tools and platforms for sharing and collaborating on spatial data and models, such as GitHub and OpenGeoHub.

Overall, the geographic complexity field is rapidly evolving, driven by advances in technology, data, and theory. These developments are leading to new insights and approaches for understanding and addressing complex spatial problems, and are likely to have significant impacts on a wide range of fields, from urban planning to environmental science.

\end{minipage}} \caption{Current trends and developments of the geographic complexity field.} \label{fig:geographic_complexity_trends} \end{figure}

\subsubsection{Complexity in biological systems}
\label{sec:biological_systems}

In this section, we explore the subfield of {\it complexity in biological systems} \citep{parisi_biology,ma2017complex}. As in the case of Sections \ref{sec:econophysics} and \ref{sec:geographic_complexity}, we start by asking ChatGPT about the purpose of this field. Figure \ref{fig:biological_systems_purpose} shows the answer to this question that says that this field investigates how complex biological systems emerge from the interactions of simpler components. Life, the emergence of ``order from order" \citep{what_is_life, kauffman,barabasi_sb}, is undoubtedly a complex process \citep{parisi_biology}. From organisms to cells, tissues, animals, societies and ecological environments, information organizes in a hierarchical way with interactions involving a wide range of scales, both in space and time \citep{siegenfeld,kauffman}. Examples include the mutation accumulation theory of aging \citep{medawar,penna,de2004penna}, neural networks, the networks of metabolic reactions inside a cell \citep{ravasz} and regulatory genes promoting proteins involved in physiological responses \citep{barabasi_grn}, the latter belonging to a new field called "Systems Biology" \citep{barabasi_sb}

In the following, we ask about the current trends and developments of this field. Figure \ref{fig:biological_systems_trends} shows the answer to this question that points the following trends {\it Multi-omics analysis} \citep{pinu2019systems,de2020computational}, {\it Single-cell analysis} \citep{spiller2010measurement}, {\it Machine learning and artificial intelligence} \citep{camacho2018next,xu2019machine}, {\it Network medicine} \citep{barabasi2007network,barabasi2011network} and {\it Synthetic biology} \citep{heinemann2006synthetic,pleiss2006promise}.

\begin{figure}[H] \centering \fbox{\begin{minipage}{36em}
\small
{\bf What is the purpose of the field that studies complexity in biological systems?
} 

\vspace{5mm}

 {\bf ChatGPT}: What is the purpose of the field that studies complexity in biological systems?

The purpose of the field that studies complexity in biological systems is to understand how complex biological systems, such as cells, organisms, and ecosystems, emerge from the interactions of simpler components, such as molecules, cells, and organisms. This field, called "systems biology," seeks to understand the underlying principles that govern the behavior of biological systems, including how they adapt to changing environments, communicate with each other, and evolve over time.

Systems biology combines experimental and theoretical approaches to study biological systems as a whole, rather than focusing on individual components in isolation. By analyzing large datasets and

\end{minipage}} \caption{The purpose of the field that studies complexity in biological systems.} \label{fig:biological_systems_purpose} \end{figure}

\begin{figure}[H] \centering \fbox{\begin{minipage}{36em}
\small
{\bf Which are the current trends and developments in the field that studies complexity in biological systems?

} 

\vspace{5mm}

 {\bf ChatGPT}: The field that studies complexity in biological systems, or systems biology, is constantly evolving and there are several current trends and developments. Here are a few examples:

    Multi-omics analysis: One of the most important trends in systems biology is the integration of multiple "omics" technologies, such as genomics, transcriptomics, proteomics, and metabolomics, to gain a more comprehensive understanding of biological systems. By combining data from these different omics platforms, researchers can identify novel regulatory networks and interactions that would be difficult to detect with a single approach.

    Single-cell analysis: Advances in single-cell technologies, such as single-cell RNA sequencing, are enabling researchers to study biological systems at an unprecedented level of detail. By analyzing gene expression and other molecular features of individual cells, systems biologists can gain insights into cellular heterogeneity, cell signaling, and the dynamics of biological systems.

    Machine learning and artificial intelligence: With the increasing availability of large-scale data sets, machine learning and artificial intelligence (AI) methods are becoming increasingly important in systems biology. These approaches can help identify patterns and relationships in complex data sets, and can be used to build predictive models of biological systems.

    Network medicine: Network medicine is a relatively new field that aims to integrate systems biology approaches with clinical data to develop more personalized approaches to diagnosis, treatment, and prevention of disease. By analyzing the interactions between genes, proteins, and other molecules in disease networks, network medicine researchers can identify novel drug targets and biomarkers for disease.

    Synthetic biology: Synthetic biology involves designing and constructing novel biological systems or modifying existing ones for specific applications. Systems biology approaches are critical for understanding and optimizing synthetic biological systems, and advances in synthetic biology are providing new tools for studying and manipulating biological systems.

\end{minipage}} \caption{Current trends and developments of the field that studies complexity in biological systems.} \label{fig:biological_systems_trends} \end{figure}

\subsection{Critiques}
\label{sec:research_complex_systems_critique}

In this subsection, as a proxy to the community, we turn to ChatGPT to explore the criticisms that the complex systems field has faced, following our earlier attempt to provide an overview of the field in Section \ref{sec:complex_systems_overview}. Figure \ref{fig:critiques_complex_systems} shows a very relevant spectrum of the critiques received by the field: {\it lack of rigorous mathematical foundations}, {\it lack of experimental validation}, {\it overreliance on reductionism}, {\it Lack of clarity on what constitutes a complex system} and {\it lack of progress in solving real-world problems}. All these critiques deserve special attention, since they are recurrent in the field. 

It is unfair to say that the field lacks mathematical rigor, since the lack of rigor is not widespread in the field. While there are some areas of the field that depend strongly on simulations, there are other areas that the mathematical rigor dominates. Among the areas with more rigor, we may cite the fields of chaos theory and information theory, with an intermediate level of rigor, we may cite the fields of complex networks and cellular automatas, and the field that depends mostly on simulations are the agent-based models (and less on mathematics). However, even if we consider the agent-based models, there are many attempts to become them more rigorous. For instance, while Challet and other physicists introduced the {\it minority game}\footnote{A minority game is a simple computational model for the collective behavior of agents where they compete through adaptation for scarce resources. It has been applied to investigate relevant issues, such as the dynamics of  financial markets \citep{jeffries2003financial,challet2004minority},  the effects of information in transportation networks \citep{kitamura2007can,BOUMAN2016capacity}, the concept of intelligence \citep{mello2008minority}, the dynamics of auctions \citep{lustosa2010constrained} and strategies in deregulated markets \citep{capodieci2016adaptive}.   } using computer simulations \citep{CHALLET1997,Savit1999} to study the El-Farol Bar problem \citep{arthur1994inductive},  there is today a mathematical basis for this model \citep{coolen2005mathematical,Ortisi2011}. Furthermore, in general in the field of agent based models, the works \cite{lorscheid2012opening} and \cite{klingert2012effectively} use the statistics of experiment design to make the results of agent-based models more credible.

The lack of experimental validation is not also widespread in the field either. In particular, we should mention the list of practical laboratory experiments presented in Section \ref{sec:teaching_lab} and many others that may be conceived following the same lines. Furthermore, a big fraction of the literature of complex systems depend on data and its consequent validation. Furthermore, even in the case of the agent based models, as we mention in Section \ref{sec:econophysics}, there are today some attempts to their validation.

While there is no universal agreement on what constitutes a complex system, current accepted definitions are highly consistent with each other. A useful resource for exploring these definitions is the {\it Journal of Physics: complexity} \citep{Bianconi_2023}, whose editors have independently provided a range of self-concordant definitions that align closely with the one presented in Section \ref{sec:definition_complex_systems}.

The lack of progress in solving real-world problems in our opinion is the less relevant critique, since models that come from the field of complex systems have dealt with many real world problems in the fields of traffic management \citep{COOK2015,cook2016complexity}, climate modeling \citep{nowack2020causal},  epidemiology \citep{Galea2009,Pastor-Satorras2015epidemic}, power grid management \citep{mei2011power,yang2017small}, ecological modeling \citep{pascual2006ecological,Anand2010}, information diffusion \citep{lind2007spreading,onnela2007structure,mello2009measuring,vespignani2012modelling,zhang2016dynamics,lengyel2020role}, political science \citep{fowler2006connecting,fowler2006legislative,koger2009partisan,fowler2011causality,nery2022co} and marketing \citep{aral2009,aral2012}. While many of these approaches depend on the structure of a complex network to model the interaction of the actors of the system, we may also find specific applications of fractals, such as cancer diagnosis and therapeutic management \citep{lennon2015lung} and the characterization of soil structure \citep{anderson1997applications}.

\begin{figure}[H] \centering \fbox{\begin{minipage}{36em}
\small
{\bf What are the criticisms about the area of complex systems in physics from the physicists of other areas?
} 

\vspace{5mm}

 {\bf ChatGPT}: There are several criticisms of complex systems in physics from physicists of other areas, including:

Lack of rigorous mathematical foundations: Some physicists criticize complex systems research for being too reliant on computer simulations and not having a solid mathematical foundation. They argue that without a firm mathematical basis, the results of complex systems research cannot be trusted.

Lack of experimental validation: Another criticism is that much of the research in complex systems has not been experimentally validated. Critics argue that without empirical evidence, the findings of complex systems research are little more than speculation.

Overreliance on reductionism: Critics argue that complex systems research is too reductionist, meaning it attempts to explain complex phenomena by breaking them down into simpler parts. They argue that this approach is limited and fails to capture the true complexity of the systems being studied.

Lack of clarity on what constitutes a complex system: Some physicists argue that there is no clear definition of what constitutes a complex system. They argue that this lack of clarity makes it difficult to apply complex systems research to real-world problems.

Lack of progress in solving real-world problems: Critics argue that despite decades of research in complex systems, little progress has been made in using this approach to solve real-world problems. They argue that the field is too theoretical and lacks practical applications.

It should be noted that not all physicists hold these criticisms, and many see the study of complex systems as a valuable and promising area of research. However, these criticisms do exist and are worth considering in any discussion of complex systems research.

\end{minipage}} 
\caption{Critiques to the field of complex systems.} 
\label{fig:critiques_complex_systems} 
\end{figure}

\subsection{Chalenges}
\label{sec:research_challenges}

In this subsection, we use ChatGPT to explore the challenges faced by the complex systems field. This is in addition to our earlier overview of the field in Subsection \ref{sec:complex_systems_overview} and our examination of criticisms of the field in Subsection \ref{sec:research_complex_systems_critique}. 

Thus, Figure \ref{fig:challenges_complex_systems} presents the potential challenges of the field of complex systems in the next few years. We may summarize the challenges in five topics: {\it dealing with big data}, {\it developing more accurate models}, {\it interdisciplinary collaboration}, {\it ethical issues} and {\it bridging theory and practice}. While these topics will undoubtedly be important for complex systems research in the years to come, it is difficult to determine their relative significance. 

Dealing with big data is actually a present issue and it is also possibly a challenge in the future. We have never produced and recorded so much data as we do today.  There are actually many specific challenges associated with the management of big data, such as data security and privacy, data integration, data quality, and cost of storing \citep{almeida2013main,chen2014,yin2015big}.   

In a world with increasing complexity, the need for more accurate models is a fundamental issue. Furthermore, the development of more accurate models can be attributed to the advancements in machine and deep learning literature, which have pushed the boundaries of predictability. 

Interdisciplinary collaboration is a trend today as we see in Subsection \ref{sec:complex_systems_overview} and it is also a challenge in the next few years. Interdisciplinary collaboration may be an important step for the acceptance of complex systems methods in other fields as we mention in Section \ref{sec:econophysics}.

Ethical issue is an important dimension in all fields that deal with data and it has been considered seriously by researchers that develop artificial inteligence and machine learning models \citep{Allen2006,Moor2006}. 
In addition, using the predictions of these models for public policy implementation requires caution, as it may have unintended negative effects on individuals or groups. Therefore, it is crucial to consider the potential consequences before making any decisions. In addition to the concerns surrounding the use of private data mentioned above, if we use the predictions of these models for the implementation of public policies \citep{kleinberg2015prediction}, we need to be careful about the consequences of these uses in order to not prejudice anyone. 

We may associate the challenging of Bridging theory and practice  with the critique to the field that complex systems do not solve real world problems as presented in Section \ref{sec:research_complex_systems_critique}. As we explore there, although we do not believe this is true, we also believe that in the next few years, the complex systems field will provide more solutions for real world problems.

\begin{figure}[H] \centering \fbox{\begin{minipage}{36em}
\small
{\bf What are the key challenges that researchers and practitioners in the field of complex systems are likely to face in the next few years?
} 

\vspace{5mm}

 {\bf ChatGPT}: As a language model, I cannot predict the future with certainty, but based on current trends and challenges in the field of complex systems, some of the key challenges that researchers and practitioners are likely to face in the next few years are:

Dealing with big data: Complex systems are often characterized by large amounts of data generated from various sources such as social media, sensors, and simulations. Analyzing and understanding these large datasets can be challenging, and new techniques and tools will be required to handle and make sense of these data.

Developing more accurate models: Complex systems are inherently unpredictable, and developing accurate models that capture the dynamics and behavior of these systems is a significant challenge. Improving model accuracy and developing new modeling techniques will be essential for making predictions and decisions based on complex systems.

Interdisciplinary collaboration: Complex systems are multidisciplinary, and understanding them requires knowledge from various fields such as physics, mathematics, computer science, biology, and social sciences. Collaborating across disciplines can be challenging, and developing effective communication and collaboration mechanisms will be necessary to advance the field.

Addressing ethical issues: Complex systems can have far-reaching effects on society, and as such, there is a need to address ethical issues such as data privacy, fairness, and accountability. Researchers and practitioners will need to develop ethical frameworks to ensure that the development and deployment of complex systems are guided by ethical principles.

Bridging theory and practice: While the field of complex systems has made significant advances in theory and modeling, there is a need to bridge the gap between theory and practice. Developing practical applications of complex systems theory and models will be essential for addressing real-world problems such as climate change, pandemics, and social inequalities.

Overall, the challenges facing researchers and practitioners in the field of complex systems are complex and multifaceted. However, by developing new techniques, collaborating across disciplines, addressing ethical issues, and bridging theory and practice, the field can continue to advance and make significant contributions to society.

\end{minipage}} 
\caption{Challenges of the field of complex systems.} \label{fig:challenges_complex_systems} \end{figure}

\section{Final remarks}

In this study, we have used ChatGPT as a representation of the community's view on the complex systems field. Through our questioning of ChatGPT, we have generated an overview of the field that encompasses both teaching and research topics. During these discussions, we supplemented the ``community" perspective with our own insights on the field. Our contributions have not only added new ingredients to the ``community"s ideas, but but also provided sources to support the community's views and also our own views.

ChatGPT can be a helpful tool for understanding the community's ideas on teaching and research because of its strong summarization skills. However, we should use it carefully because if the data it uses to learn about the community has biases, ChatGPT may also learn and reproduce those biases. Furthermore, the popularity bias may also affect ChatGPT's answers paying attention to the most popular issues instead of necessarily the most important ones.

\section{Acknowledgment} 

DOC (grant number 302629/2019-0) and NC (grant number 310893/2020-8) thank CNPq for partial financial support. NC also thanks FAPERJ (grant number 203.217/2017). We want to reinforce that ChatGPT has written its all answers.

\end{document}